\documentclass[sigconf]{acmart}

\usepackage{type1cm}     
\usepackage{graphicx}     
\usepackage{xspace}     
\usepackage{balance}     
\usepackage{multirow}     
\usepackage[font={bf}, tableposition=top]{caption}     
\usepackage{bold-extra}     
\usepackage{siunitx}          
\usepackage[vlined,linesnumbered,ruled,noend]{algorithm2e}     
\usepackage{booktabs}     
\usepackage{microtype}    
\usepackage{units}     
\usepackage{mathtools}     
\usepackage[hide]{chato-notes}
\usepackage{relsize}
\usepackage{caption}
\usepackage[]{inputenc}
\usepackage{xfrac}
\usepackage{balance}
\RequirePackage{graphicx,color}
\usepackage[font={small}]{subfig} 
\usepackage{pifont}
\usepackage{footnote} 
\makesavenoteenv{table}

\newcommand{\ourtitle}{Political Discourse on Social Media:\\ Echo Chambers, Gatekeepers, and the Price of Bipartisanship}

\definecolor{MyWhite}{rgb}{1.00,1.00,1.00}
\newcommand\txtwhite[1]{{\color{MyWhite}{#1}}}

\newcommand{\spara}[1]{\smallskip\noindent\textbf{#1}}

\newcommand{\cmark}{\ding{51}\xspace}
\newcommand{\xmark}{\ding{55}\xspace}

\newcommand{\controversial}{\textsc{Political}\xspace}
\newcommand{\largetw}{\texttt{large}\xspace}
\newcommand{\guncontrol}{\texttt{guncontrol}\xspace}
\newcommand{\obamacare}{\texttt{obamacare}\xspace}
\newcommand{\abortion}{\texttt{abortion}\xspace}
\newcommand{\election}{\texttt{combined}\xspace}

\newcommand{\noncontroversial}{\textsc{Non-Political}\xspace}
\newcommand{\ff}{\texttt{ff}\xspace}
\newcommand{\love}{\texttt{love}\xspace}
\newcommand{\gameofthrones}{\texttt{gameofthrones}\xspace}
\newcommand{\tbt}{\texttt{tbt}\xspace}
\newcommand{\foodporn}{\texttt{foodporn}\xspace}

\newcommand{\leaning}{\emph{political leaning}\xspace}
\newcommand{\production}{\emph{production polarity}\xspace}

\newcommand{\consumption}{\emph{consumption polarity}\xspace}

\newcommand{\rtrate}{\emph{retweet rate}\xspace}
\newcommand{\rtvol}{\emph{retweet volume}\xspace}
\newcommand{\favrate}{\emph{favorite rate}\xspace}
\newcommand{\favvol}{\emph{favorite volume}\xspace}

\newcommand{\deltapartisan}{{$\delta$}-partisan\xspace}
\newcommand{\deltabipartisan}{{$\delta$}-bipartisan\xspace}
\newcommand{\deltaconsumer}{{$\delta$}-consumer\xspace}
\newcommand{\deltakeeper}{{$\delta$}-gatekeeper\xspace}

\newenvironment {squishlist}
{\begin{list}{$\bullet$}
  { \setlength{\itemsep}{0pt}
     \setlength{\parsep}{3pt}
     \setlength{\topsep}{3pt}
     \setlength{\partopsep}{0pt}
     \setlength{\leftmargin}{1.5em}
     \setlength{\labelwidth}{1em}
     \setlength{\labelsep}{0.5em} } }
{\end{list}}

\title{\ourtitle}

\author{Kiran Garimella}
\affiliation{%
  \institution{Aalto University}
  \city{Espoo}
  \country{Finland} 
}
\email{kiran.garimella@aalto.fi}

\author{Gianmarco De~Francisci~Morales}
\affiliation{%
  \institution{Qatar Computing Research Institute}
  \city{Doha} 
  \country{Qatar} 
}
\email{gdfm@acm.org}

\author{Aristides Gionis}
\affiliation{%
  \institution{Aalto University}
  \city{Espoo} 
  \country{Finland} 
}
\email{aristides.gionis@aalto.fi}

\author{Michael Mathioudakis}
\affiliation{%
  \institution{University of Helsinki}
  \city{Helsinki} 
  \country{Finland} 
}
\email{michael.mathioudakis@helsinki.fi}

\begin{abstract}
Echo chambers, i.e., situations where one is exposed only to opinions that agree with their own, 
are an increasing concern for the political discourse in many democratic countries. 
This paper studies the phenomenon of political echo chambers on social media.
We identify the two components in the phenomenon: the opinion that is shared, and 
the ``chamber'' (i.e., the social network) that allows the opinion to ``echo'' 
(i.e., be re-shared in the network) -- and 
examine closely at how these two components interact.
We define a production and consumption measure for social-media users, 
which captures the political leaning of the content shared and received by them.
By comparing the two, we find that Twitter users are, to a large degree, 
exposed to political opinions that agree with their own.
We also find that users who try to bridge the echo chambers, by sharing content with diverse leaning, 
have to pay a ``price of bipartisanship'' in terms of their network centrality and content appreciation.
In addition, we study the role of ``gatekeepers,'' 
users who consume content with diverse leaning but produce partisan content (with a single-sided leaning), 
in the formation of echo chambers.
Finally, we apply these findings to the task of predicting partisans and gatekeepers from social and content features.
While partisan users turn out relatively easy to identify, gatekeepers prove to be more challenging.
\end{abstract}

\begin{document}

\copyrightyear{2018}
\acmYear{2018}
\setcopyright{iw3c2w3}
\acmConference[WWW 2018]{The 2018 Web Conference}{April 23--27, 2018}{Lyon, France}
\acmBooktitle{WWW 2018: The 2018 Web Conference, April 23--27, 2018, Lyon, France}
\acmPrice{}
\acmDOI{10.1145/3178876.3186139}
\acmISBN{978-1-4503-5639-8/18/04}

\fancyhead{}
\maketitle

\renewcommand{\shortauthors}{K. Garimella, G. De Francisci Morales, A. Gionis, M. Mathioudakis}
\section{Introduction}
\label{sec:intro}

\emph{Echo chambers} have emerged as an issue of concern in the political discourse of democratic countries. 
There is growing concern that, as citizens become more polarized about political issues, they do not hear the arguments of the opposite side, 
but are rather surrounded by people and news sources who express only opinions they agree with.
It is telling that 
Facebook and ex-U.S. Presidents have recently voiced such concerns.\footnote{E.g., Obama foundation's attempt to address the issue of echo chambers.~\url{https://www.engadget.com/2017/07/05/obama-foundation-social-media-echo-chambers}}
If echo chambers exist, then they might hamper the deliberative process in democracy
\citep{sunstein2009republic}.

In this paper, we study the degree to which echo chambers exist in political discourse on Twitter, and how they are structured.
We approach the study in terms of two components: the opinion that is shared by a user, and 
the ``chamber'', i.e., the social network around the user, which allows the opinion to ``echo'' back to the user as it is also shared by others.
The opinion corresponds to \emph{content} items shared by users, while the underlying social \emph{network} is what allows their propagation.
We say that an echo chamber exists if \emph{the political leaning of the content that users receive from the network agrees with that of the content they share}.

As there is no consensus on a formal definition in the literature, we opt for this definition, which is general enough and reasonably captures the essence of the phenomenon.
There are, however, a few previous works that have studied echo chambers under different perspectives.
For instance, previous works have focused either on the differences between the content shared and read by partisans of different sides~\citep{garrett2009echo, gilbert2009blogs, an2014partisan, quattrociocchi2016echo}; the social network structure~\citep{gromping2014echo}; or the structure of user interactions, such as blog linking~\cite{adamic2005political} and retweets~\citep{conover2011political, garimella2016quantifying}.
We adopt a definition which is broader in terms of \emph{content} it is based on 
(it considers all content shared and produced, not only content pertaining to specific types of interactions, e.g., retweets), and which is defined jointly on \emph{content} and \emph{network}.


Specifically, we define production and consumption measures for social media users based on the political leaning of the content \emph{shared with} and \emph{received from} their network.
We apply them to several datasets from Twitter, including a large one consisting of over 2.5 billion tweets, which captures 8 years worth of exchanges between politically-savvy users.
Our findings indicate there is large correlation between the leaning of content produced and consumed: \emph{echo chambers are prevalent on Twitter}.

We then proceed to analyze \emph{partisan} users, 
who produce content with predominantly one-sided leaning,\footnote{We use ``leaning'' as a score that quantifies alignment with one political side. Similar terms in the literature include ``ideology,'' ``polarity,'' or ``ideological stance.''}
and \emph{bipartisan} users, which instead produce content with both leanings.
Our analysis indicates that partisan users enjoy a higher ``appreciation'' as measured by both network and content features.
This finding hints at the existence of a \emph{``price of bipartisanship,''} required to be paid by users who try to bridge echo chambers.


Moreover, we take a closer look on \emph{gatekeeper} users, who consume content of both leanings, but produce content of a single-sided leaning.
These users are \emph{border spanners} in terms of location in the social network, who remain aware of the positions of both sides, but align their content with one side.
They are a small group, which enjoy higher than average network centrality, while not being very embedded in their community.

Finally, we use these findings for predicting \emph{partisan} and \emph{gatekeeper} users by using features from the content they produce and from their social network.
While partisan users are relatively easy to identify, gatekeepers prove to be more challenging.

Our study opens the road for further investigation of the echo chamber phenomenon.
While establishing the existence of political echo chambers on Twitter, based on a broad definition and measurements over a large volume of data, it also invites a more nuanced analysis of such phenomenon -- one that, instead of categorizing users in terms of partisanship, takes into account a variety of user attitudes (e.g., partisans, gatekeepers, and bipartisans).
Such analysis might be crucial to understand how to nudge users towards consuming content that challenges their opinion and thus bridge echo chambers.
Furthermore, our study shows the interdependence between content production \& consumption and network properties in the context of echo chambers. 
This finding could help us in revisiting existing models
for the dynamics of opinion formation and polarization on social networks~\cite{dellaposta2015liberals,qiu2017limited} 
that take into account not only the opinion (content) spread over the social network, but also its impact of structure of the network itself.
\section{Related work}
\label{sec:related}

\spara{Echo chambers.}
The term refers to situations where people ``hear their own voice'' 
--- or, particularly in the context of social media, 
situations where users consume content that expresses the same point of view 
that users themselves hold or express.
Echo chambers have been shown to exist in various forms of online media such as 
blogs~\cite{wallsten2005political,gilbert2009blogs}, forums~\cite{edwards2013participants}, and 
social-media sites~\cite{gromping2014echo,barbera2015tweeting,quattrociocchi2016echo}.

Previous studies have tried to quantify the extent to which echo chambers exist online.
For example, in the context of blogs, \citet{gilbert2009blogs} study the comments on a set of political
blogs and find that comments disproportionately agree with the author of the blog post.
Similar findings were reported by \citet{lawrence2010self}, who found that partisan bloggers engage with blogs of a narrow spectrum of political views, which agreed with their own.
In the context of Twitter, \citet{an2014sharing} analyzed the activity of users who engage with political news, and found that ``90\% of the users [directly follow] news media of only one political leaning'', while ``their friends' retweets lead them to diversify their news consumption''.

In the context of Facebook, \citet{bakshy2015exposure} measure the degree to which users with declared political affiliations consume cross-cutting content, i.e., content predominantly posted by users of opposing political affiliation.
Content consumption is studied at three levels:
($i$)~\emph{potential exposure}, which includes all content shared by the friends of a user;
($ii$)~\emph{exposure}, which includes all content appearing in the feed of a a user;
and ($iii$)~\emph{engagement}, which includes all content that a user clicks.
The study finds that, even though users are exposed to a significant amount of cross-cutting content, 
users opt to engage with 
less cross-cutting content,
a behavior compatible with the theory of {\em biased assimilation}~\cite{lord1979biased}.
In our work, we study content consumption at the level of \emph{potential exposure}, as a study at the remaining two levels requires access to data that is not publicly available.
%
%
However, there is no consistent definition of what an echo chamber represents in the literature. 
The studies presented above measure different aspects of an echo chamber, and 
focus \emph{either} on the content~\cite{bakshy2015exposure,gilbert2009blogs,lawrence2010self} 
\emph{or} the network~\cite{an2014sharing,adamic2005political} aspect.

In this paper, we propose measures to identify the existence of an echo chamber by using 
\emph{both} the content being read/shared \emph{and} the network that enables the content to propagate.
Unlike many previous works that focus on measuring only content consumption to quantify the echo-chamber effects, we study content consumption and production jointly at the level of individual users, and 
examine how different content profiles correlate with the network position of users.
Though we are not the first to study echo chambers on Twitter, to the best of our knowledge, this is the 
first study to jointly use content and network to characterize echo chambers.


\spara{Psychological and algorithmic mechanisms.}
\emph{Selective exposure theory}~\cite{frey1986recent} --- 
which proposes the concepts of \emph{selective exposure}, 
\emph{selective perception}, and 
\emph{selective retention} --- 
is the tendency of individuals to favor information that aligns with their pre-existing views 
while avoiding contradictory information.
\emph{Biased assimilation}~\cite{lord1979biased}, on the other hand, 
is a related phenomenon, where an individual gets exposed to information from all sides, 
but has the tendency to interpret information in a way that supports a pre-existing opinion.
All these psychological mechanisms, 
together with other biases, such as, 
\emph{algorithmic filtering} and \emph{personalization}~\cite{bozdag2013bias}, 
are connected to the phenomenon of echo chambers.
Understanding how all these phenomena interact with each other and the precise
causality relations is beyond the scope of this paper.

%

\spara{Relationship between node and network properties.}
One of our objectives is to understand the relationship between node properties 
(user consumption and production) and network properties 
(e.g., PageRank and clustering coefficient).

Homophily is a central notion in the study of social networks.
Given a network and a node feature, 
homophily refers to the phenomenon where neighboring nodes in the network 
tend to present similar values of the given feature.
Several studies have provided evidence of homophily in social networks~\citep{mcpherson2001birds}.
For example, in the context of Twitter, 
clusters in retweet networks have been found to correlate 
with the political ideologies of Twitter users~\citep{barbera2015tweeting, conover2011political, garimella2016quantifying}.
The notion of echo chambers we study here can be seen as form of 
homophily, where we consider the political leaning of content shared by users as a feature.


\spara{Price of bipartisanship.}
\citet{hetherington2001resurgent} argues that political parties have increased their prominence in the masses by being more partisan.
\citet{prior2013media} analyzes the role of partisan media to answer the question: 
``has partisan media created political polarization and led the American public to support more partisan policies and candidates?'' 
They find no evidence to support that claim.
Conversely, \citet{dellavigna2007fox} show that Fox News, being partisan and biased, could affect senate vote share and voter turnout.
They estimate that Fox News convinced 3 to 8 percent of its viewers to vote Republican.

In this paper, we study the price of being bipartisan, for the first time on social networks.
We show that producing content that expresses opinions aligned with both sides of the political divide, 
has a cost in terms of centrality in the network and content-engagement rate.

\spara{Gatekeeping.}
Gatekeeping is a term commonly used in communication studies to refer to news media sources that act as filters of information~\cite{lewin1943forces}.
\citet{barzilai2009gatekeeping} propose a model based on network theory for gatekeeping which generalizes the concept of gatekeeping for the Internet and applies to all information types (not just news).
Several studies have looked at gatekeeping practices on Twitter~\cite{kwon2012audience,xu2014talking} and conclude that unlike in traditional media, any common user can become a gatekeeper on social media.
The definition of these gatekeepers on social media also differs from the traditional gatekeepers in media organizations, due to the alternative information pathways available to social media users.

In our case, we define gatekeepers as users who receive content from both political leanings, but only produce content from a single leaning, thus ``filtering'' information from one side.
To the best of our knowledge, 
this is the first paper to study the role of gatekeepers of information within echo chambers.

\section{Data}
\label{sec:datasets}



We use a collection of ten different datasets from Twitter, 
each of which contains a set of tweets on a given topic of discussion.
The datasets span a long period of time and cover a wide range of users and topics, described below.
The collection is partitioned into two groups,
\controversial and \noncontroversial,
depending on whether the topic of discussion is politically contentious or not.
Moreover, in addition to tweets, for each dataset, we build 
a network that represents the social connections among users.
The size of each dataset in terms of number of tweets and number of distinct users is shown in Table~\ref{tab:datasets}.
For all the datasets, we perform simple checks to remove bots, using minimum and maximum thresholds for number of tweets per day, followers, friends, and ensure that the account is at least one year old at the time of data collection.
More details about the datasets are given below. 


\spara{\controversial}.
Five of the ten Twitter datasets are relevant to well-known political controversies.
Three of these datasets, namely \guncontrol, \obamacare, and \abortion, discuss a specific topic.
Each dataset is built by collecting tweets posted during specific events 
that led to an increased interest in these topics (see Table~\ref{tab:datasets}).
%
Using the Archive Twitter Stream grab,\footnote{\url{https://archive.org/details/twitterstream}}
we select tweets that contain keywords pertaining to each topic that were posted in a time period of one week around the event (3 days before and 3 days after the event).\footnote{We use the keyword lists proposed by~\citet{lu2015biaswatch}.}
%
To focus on users who are actively engaged in the discussion of each topic,
we identify the subset of users who have at least $5$ tweets about the topic during this time window.
We collect all the tweets posted by these users via Twitter's REST API.\footnote{\url{https://developer.twitter.com/en/docs/tweets/timelines/overview}}
The datasets are obtained by~\citet{garimella2017long} and have already been validated in previous work~\cite{garimella2017effect}. 


A fourth dataset, named \election, is collected in a similar fashion, 
except that it contains tweets of users who were active during the U.S. presidential election results of 2016 (November 6--12, 2016),
and who tweeted at least $5$ times about any of the three controversial topics \guncontrol, \obamacare, and \abortion.
We also collect all tweets of these users via Twitter's REST API.

Finally, the fifth dataset, named \largetw, is a large dataset containing over 2.5 billion tweets from politically active users spanning a period of almost 8 years (2009-2016). 
Specifically, the dataset consists of all tweets generated by users who retweeted a presidential or vice-presidential candidate from 2008-2016 in the U.S. at least $5$ times. 
The dataset has been used in previous work~\cite{garimella2017long}; we refer to the original paper for more details.

\spara{\noncontroversial}. To have a baseline for our measurements over
the \controversial\ datasets, we also use five datasets that correspond to
non-political topics, in particular:
\tbt (``throwback Thursday''), 
\ff (``follow Friday''), 
\gameofthrones, 
\love, 
and \foodporn.
Each of these topics is associated with a particular hashtag (e.g., \#tbt for \tbt).
The datasets are built as follows.
First, we parse the tweets in the Internet Archive collection and select tweets that contain the corresponding hashtag for each topic during the month of June 2016.
Second, we filter out users who have less than $5$ tweets.
Third, we obtain all tweets generated by these users.
The resulting set of tweets for each topic constitutes one dataset.

\spara{Network}. For each dataset,
we build the directed ``follow'' graph among users: 
an edge $(u\rightarrow v)$ indicates that user $u$ follows user $v$.

\begin{table}[t]
\centering
\footnotesize
\caption{Description of the datasets.}
\label{tab:datasets}
\begin{tabular}{l r r p{3cm}}
\toprule
Topic & \#Tweets & \#Users & Event \\
\midrule
\guncontrol & 19M & \num{7506} & Democrat filibuster for guncontrol reforms (June 12--18, 2016)\footnote{\url{https://en.wikipedia.org/wiki/Chris_Murphy_gun_control_filibuster}} \\
\obamacare & 39M & \num{8773} & Obamacare subsidies preserved in U.S. supreme court ruling (June 22--29, 2015)\footnote{\url{http://www.bbc.com/news/world-us-canada-33269991}} \\
\abortion & 34M & \num{3995} & Supreme court strikes down Texas abortion restrictions (June 27--July 3, 2016)\footnote{\url{https://www.nytimes.com/2016/06/28/us/supreme-court-texas-abortion.html}} \\
\election & 19M & \num{6391} & 2016 US election result night (Nov 6--12, 2016) \\
\largetw & 2.6B & \num{676996} & Tweets from users retweeting a U.S. presidential/vice presidential candidate (from~\cite{garimella2017long}, 2009--2016) \\
\midrule
\ff & 4M & \num{3204} & \multirow{5}{*}{filtering for these hashtags}\\
\gameofthrones & 5M & \num{2159} & \\
\love & 3M & \num{2940} & \\
\tbt & 28M & \num{12778} & \\
\foodporn & 8M & \num{3904} & \\
\bottomrule
\end{tabular}
\vspace{-\baselineskip}
\end{table}

\spara{Political leaning scores (source polarity).} 
Our analysis relies on characterizing the political leaning of the content consumed and produced by each user. 
Obtaining a characterization of political leaning for short text snippets, such as tweets, is a very challenging problem, in general. 
To confront this challenge, we use a ground truth of political leaning scores of various news organizations with a presence on social media obtained from~\citet{bakshy2015exposure}.
Specifically, the data contains a score of political leaning for $500$ news domains (e.g., \emph{nytimes.com}) that are most shared on Facebook.
The score takes values between $0$ and $1$ and expresses the fraction of Facebook users who visit these pages that identify themselves as conservative on their Facebook profile.
A value close to 1 (0) indicates that the domain has a conservative (liberal) bent in their coverage.
For a detailed description of the dataset, we refer the reader to the original publication~\cite{bakshy2015exposure}.
We remove a small number of domains that are not owned by news organizations
(e.g., \emph{wikipedia.org} or \emph{reddit.com}),
and add shortened versions of news domains to the list (e.g. \emph{fxn.ws} for \emph{foxnews.com}).
The distribution of source polarity for the 500 domains is shown in Figure~\ref{fig:content_polarity_distribution}.

\section{Measures}
\label{sec:measures}

This section describes the measures used in our analysis.
These measures aim to capture user activity from two perspectives: 
($i$)~the \emph{content} produced and consumed by a user, and 
($ii$)~the \emph{network} position of a user, including their interactions with others.

\subsection{Content}

Content is central in measuring echo chamber effects. 
In a setting where opinions are polarized between two perspectives --
in our case ``liberal'' and ``conservative'' --
we say that \emph{an echo chamber exists to the degree that users consume content that 
agrees with their expressed point of view}.
To make this definition actionable and quantify the echo chamber effect, 
we need to model the \emph{political leaning} of content \emph{produced} and \emph{consumed} by users.

For the \emph{content production} of a user $u$, 
we consider tweets posted by user $u$.
For the \emph{content consumption} of a user $u$
we consider tweets posted by users whom $u$ follows.

To quantify the political leaning of content posted on Twitter, 
we consider only messages that contain a link to an online news
organization with a known and independently derived political leaning.
In particular, we use the dataset of the political leaning scores
of news organizations described in Section~\ref{sec:datasets}.
Based on those scores, we define a polarity score for the content produced and consumed
by a user.

\spara{Production polarity}. For each user $u$ in a given dataset, 
we consider the set of tweets $P_u$ posted by $u$ that contain links to news organizations of 
known \leaning $l_n$.
We then associate each tweet $t\in P_u$ with leaning $\ell(t) = l_n$.
The \production $p(u)$ of user $u$ is then defined as the average \leaning 
over $P_u$, i.e., 
\begin{equation}
	p(u) = \frac{\sum_{t\in P_u} \ell(t)}{|P_u|}.
\end{equation}
The value of \production\ ranges between $0$ and $1$.
For users who regularly share content from liberal sources, \production\ is closer to $0$, 
while for the ones who share content from conservative sources it is closer to $1$. 

\begin{figure}
\centering
\includegraphics[width=0.45\textwidth]{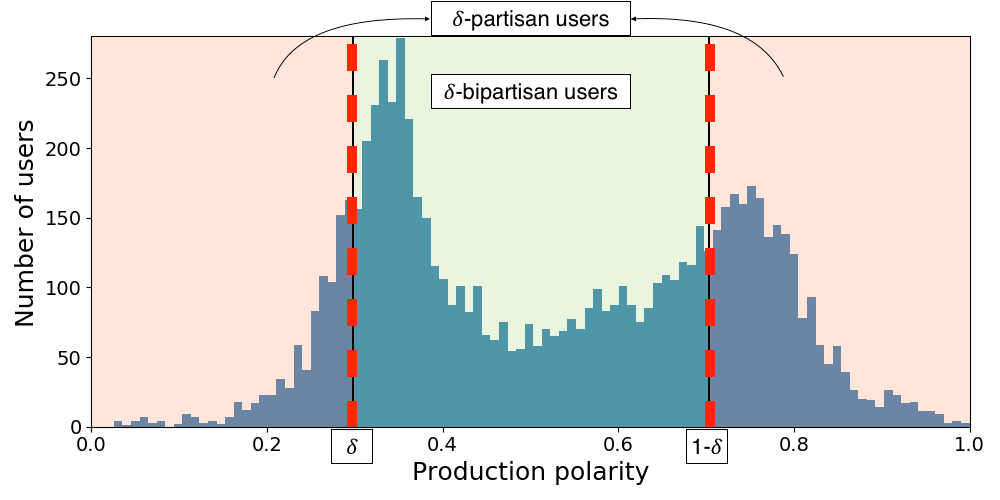}
\caption{Example showing the definition of \deltapartisan users. The dotted red lines are drawn at $\delta$ and 1-$\delta$. Users on the left of the leftmost dashed red line or right of the rightmost one are \deltapartisan.}
\label{fig:example}
\vspace{-\baselineskip}
\end{figure}

We wish to quantify the extent to which users produce one-sided content.
We say that a user is \textbf{\deltapartisan}, for some value $0\le\delta\le\frac{1}{2}$, 
if their \production is within $\delta$ from either extreme value
\begin{equation}
\min\{p(u), 1-p(u)\} \leq \delta.
\end{equation}
The smaller the value of $\delta$ the more partisan a user is.
Note also that 
if a user~$u$ is \deltapartisan then $u$ is also 
$\delta'$-partisan for $\delta<\delta'\le\frac{1}{2}$.
Users who are not \deltapartisan are called \deltabipartisan.
Intuitively, \deltapartisan users produce content only from one extreme end of the political spectrum, where as \deltabipartisan ones do not.
Figure~\ref{fig:example} shows an illustration of \deltapartisan and \deltabipartisan users.

\spara{Production variance}. Besides the average \leaning\ of produced tweets,
we also measure the {\it variance} in \leaning\ over the same set of tweets. 
The objective is to quantify the range of opinions of a user covered by the produced content.

\spara{Consumption polarity}.
Similarly to \production, 
we define \consumption based on the set of tweets $C(u)$ that
a user receives on their feed from users they follow.
We again focus on tweets that contain a link
to a news article from a domain with known source polarity.
The \consumption $c(u)$ of user $u$ is defined as the average \leaning\ of 
received tweets $C(u)$.
\begin{equation}
	c(u) = \frac{\sum_{t\in C_u} \ell(t)}{|C_u|}
\end{equation}
Values close to $0$ indicate consumption of liberal content, while values
close to $1$ indicate consumption of conservative content.
Although the definition of \consumption is based on the source polarity of tweets, 
it also takes the network structure into account and
forms the basis for the understanding of the interaction between content and network.

To quantify the extent to which users consume one-sided content, 
we say that a user is {\bf \deltaconsumer}, for some value $0\le\delta\le\frac 12$,
if their \consumption is within $\delta$ from either extreme value
\begin{equation}
\min\{c(u), 1-c(u)\} \leq \delta. 	
\end{equation}

\spara{Consumption variance}. Besides the average \leaning\ of consumed tweets,
we also measure the {\it variance} in \leaning\ over the same set of tweets.
The objective is to quantify the range of opinions of a user covered by the consumed content.

\spara{Gatekeepers}.
Gatekeepers are defined in media and communication studies as media sources that act as filters (or `gatekeepers') of information~\cite{lewin1943forces}. 
In our case, we consider consumption and production of content jointly, 
and define gatekeepers as users who consume content from both sides of the political spectrum but only produce content from one side. 
These users block or filter information from one side, and hence can be considered gatekeepers.

Formally, we say that a user $u$ is {\bf \deltakeeper} if $u$
is \deltapartisan but {\bf not} \deltaconsumer, i.e., 
\begin{equation}
\label{eq:gatekeepers}
\min\{p(u), 1-p(u)\} \leq \delta \:{\text{ and }}\: \min\{c(u), 1-c(u)\} > \delta.
\end{equation}

\subsection{Network} 

Our goal is to understand the interplay of content consumption and production
with the position of the users in the network and the global network structure. 
Thus, to add to the above measures defined using content, we define measures that
capture the position of the user in a network and their interactions
with other users.
We consider the following network measures.

\spara{User polarity}.
We adopt the latent space model proposed by~\citet{barbera2015tweeting} to estimate a \emph{user polarity} score.
This score is based on the assumption that Twitter users prefer to follow politicians whose position
on the latent ideological dimension is similar to theirs.
For the list of politicians and details on estimating the polarity, please refer to the original paper~\cite{barbera2015tweeting}.
Negative (positive) values of the user polarity scores indicate a democrat (republican) leaning and
the absolute value of the polarity indicates the degree of support to the respective party.

\spara{Network centrality}. 
We employ the well-known PageRank measure~\cite{page1999pagerank} to characterize the centrality of a node in a network.
PageRank reflects the importance of a node in the follow network, and a higher PageRank can be interpreted as a higher chance of the user to spread its content to its community.

\spara{Clustering coefficient}.
In an undirected graph, the clustering coefficient $\mathit{cc}(u)$ of a node $u$ is defined as the fraction of closed triangles in its immediate neighborhood. 
Specifically, let $d$ be the degree of node $u$, and $T$ be the number of 
closed triangles involving $u$ and two of its neighbors. 
The clustering coefficient is then defined as $\mathit{cc}(u) = \frac{2T}{d\left(d-1\right)}$. 
Note that, as the networks in our datasets are directed graphs, we consider their undirected version to compute clustering coefficients.
A high clustering coefficient for a node indicates that the ego network of the corresponding user is tightly knit, i.e., the node is embedded in a well-connected community. 

\spara{Retweet/Favorite rate}. For a given dataset, the \rtrate\ (\favrate) of a user
is the fraction of the tweets of that user that have received at least one retweet (favorite).

\spara{Retweet/Favorite volume}. For a given dataset, the \rtvol\ (\favvol) of a user
is defined as the median number of retweets (favorites) received by their tweets.
This is different from the retweet/favorite rate because it indicates the popularity of the content, where as the retweet/favorite rate captures ``acceptance'' of the user's content.

\section{Analysis}
\label{sec:experiments}

In this section, we analyze the datasets 
described in Section~\ref{sec:datasets} 
by using the measures defined in Section~\ref{sec:measures} 
in order to answer the following questions:
\begin{squishlist}
\item[(1)] Are there echo chambers or are users exposed to content that expresses opposite leaning? 
We answer these questions by examining the joint distribution of production and consumption polarities (\S\,\ref{subsec:production_consumption}).
\item[(2)] Is there an advantage in being partisan? 
We quantify advantage in terms of network centrality (PageRank) and connectivity
(clustering coefficient),
as well as in terms of content appreciation 
(number of retweets and favorited tweets) (\S\,\ref{subsec:partisans}).
\item[(3)] Who are the users who act as gatekeepers of information in the network?
We explore features of these users and examine how they differ from other users. (\S\,\ref{subsec:gatekeepers}).
\item[(4)] Can we predict if a user is a partisan or a gatekeeper, just by examining their tweets?
We build a classification model that predicts if a user is a partisan or a gatekeeper, leveraging features extracted from the above analysis (\S\,\ref{section:prediction}).
\end{squishlist}

\subsection{Echo chambers: content production and consumption}
\label{subsec:production_consumption}

\begin{figure}
\centering
\includegraphics[width=0.45\textwidth]{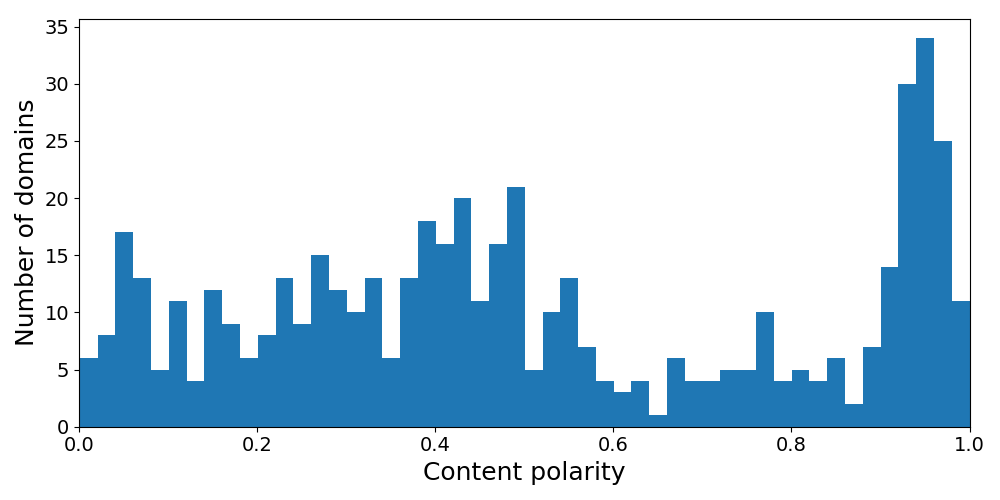}
\caption{Distribution of source polarity for the 500 news sources considered in the current work~\citep{bakshy2015exposure}.}
\label{fig:content_polarity_distribution}
\vspace{-\baselineskip}
\end{figure}

As discussed in Section~\ref{sec:measures}, the political leaning of produced and consumed content is measured based on the leaning of cited news sources.
The distribution of source polarity scores for the news sources is shown in Figure~\ref{fig:content_polarity_distribution}.
The distribution shows that there are many conservative outlets, and a sizeable number of neutral and liberal outlets.

\begin{figure*}[ht]
\centering
\begin{minipage}{.19\linewidth}
\centering
\subfloat[]{\label{}\includegraphics[width=\textwidth, height=\textwidth]{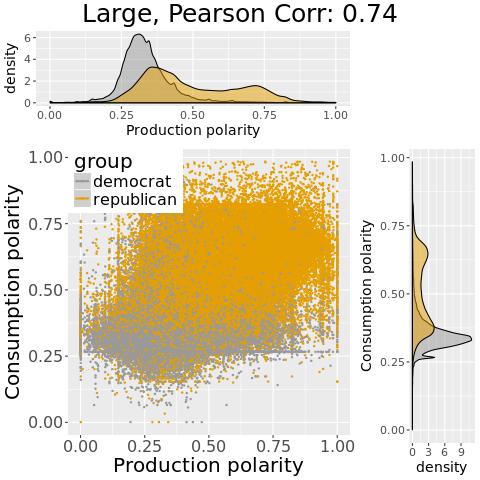}}
\end{minipage}%
\begin{minipage}{.19\linewidth}
\centering
\subfloat[]{\label{}\includegraphics[width=\textwidth, height=\textwidth]{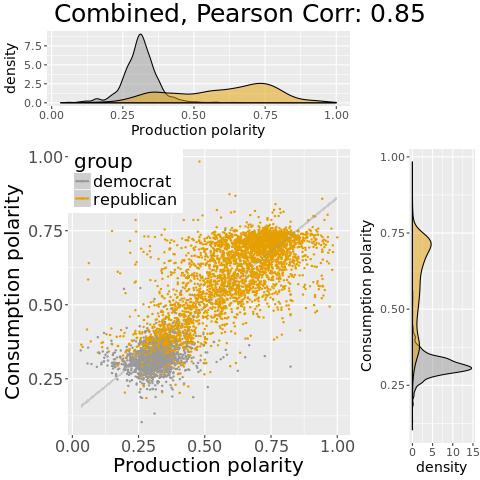}}
\end{minipage}%
\begin{minipage}{.19\linewidth}
\centering
\subfloat[]{\label{}\includegraphics[width=\textwidth, height=\textwidth]{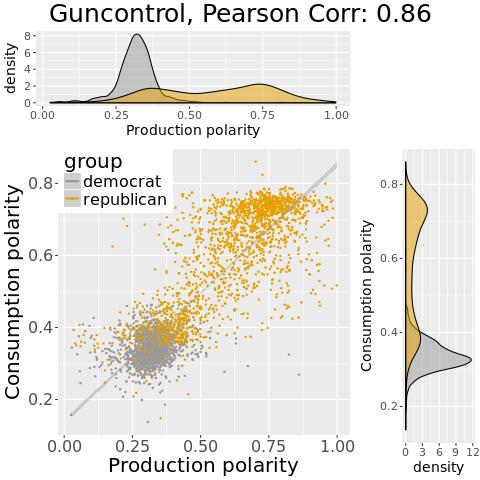}}
\end{minipage}%
\begin{minipage}{.19\linewidth}
\centering
\subfloat[]{\label{}\includegraphics[width=\textwidth, height=\textwidth]{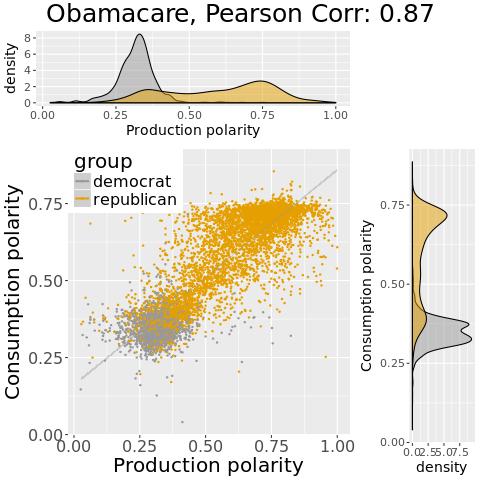}}
\end{minipage}%
\begin{minipage}{.19\linewidth}
\centering
\subfloat[]{\label{}\includegraphics[width=\textwidth, height=\textwidth]{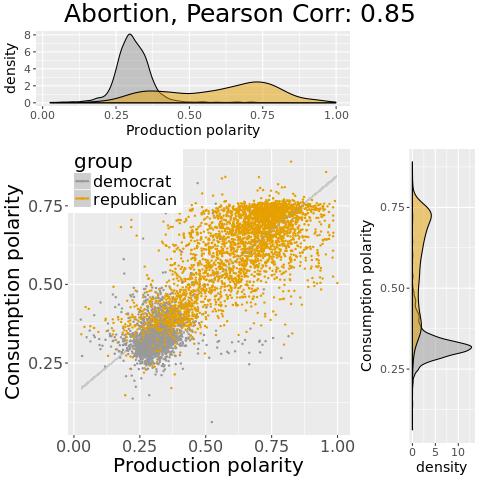}}
\end{minipage}%
\par\medskip
\begin{minipage}{.19\linewidth}
\centering
\subfloat[]{\label{}\includegraphics[width=\textwidth, height=\textwidth]{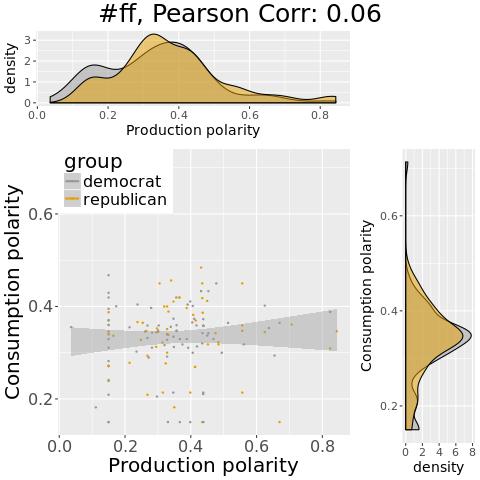}}
\end{minipage}%
\begin{minipage}{.19\linewidth}
\centering
\subfloat[]{\label{}\includegraphics[width=\textwidth, height=\textwidth]{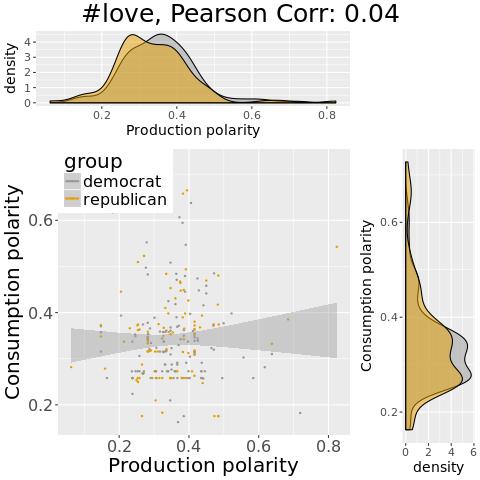}}
\end{minipage}%
\begin{minipage}{.19\linewidth}
\centering
\subfloat[]{\label{}\includegraphics[width=\textwidth, height=\textwidth]{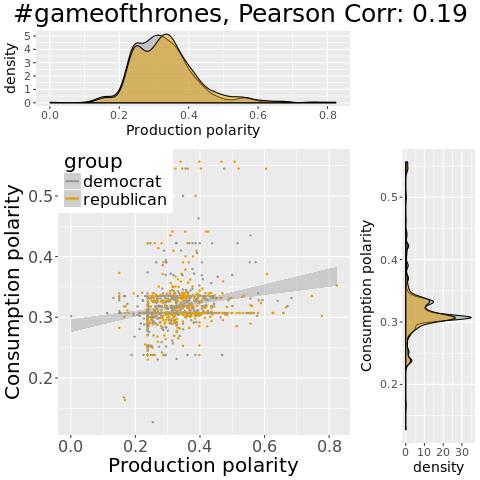}}
\end{minipage}%
\begin{minipage}{.19\linewidth}
\centering
\subfloat[]{\label{}\includegraphics[width=\textwidth, height=\textwidth]{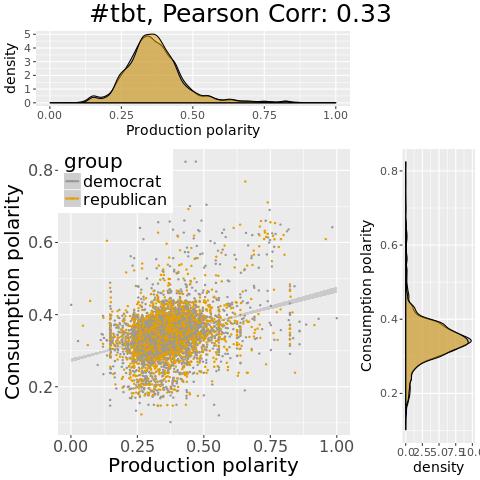}}
\end{minipage}%
\begin{minipage}{.19\linewidth}
\centering
\subfloat[]{\label{}\includegraphics[width=\textwidth, height=\textwidth]{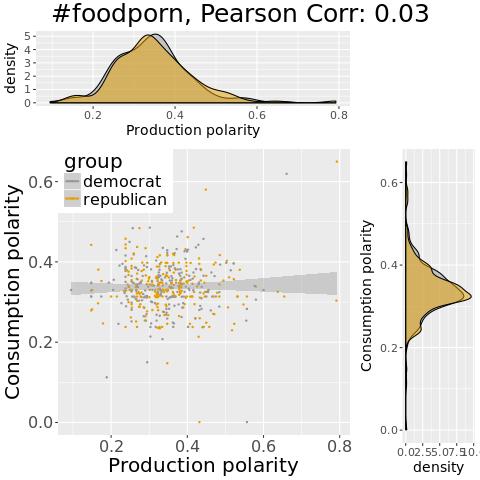}}
\end{minipage}%
\par\medskip
\caption{Distribution of production and consumption polarity, for \controversial (first row) and \noncontroversial (second row) datasets. The scatter plots display the production ($x$-axis) and consumption ($y$-axis) polarities of each user in a dataset. Colors indicate user polarity sign, following~\cite{barbera2015birds} (grey = democrat, yellow = republican). The one-dimensional plots along the axes show the distributions of the production and consumption polarities for democrats and republicans.}
\label{fig:production_consumption}
\vspace{-\baselineskip}
\end{figure*}

To explore the values of production and consumption polarities across the datasets, 
let us examine Figure~\ref{fig:production_consumption}.
The top row shows five plots for the \controversial datasets, 
and the bottom row for the \noncontroversial ones.
Each plot contains three subplots: a two-dimensional scatter-plot in the center and two one-dimensional subplots along the two axes of the scatter-plot.

The distribution of production and consumption polarities of users in the various datasets is shown in the scatter plots of Figure~\ref{fig:production_consumption}.
Each point in the scatter-plot corresponds to a user.
Recall that lower polarities indicate liberal users,
and higher polarities indicate conservative ones.
The color of each point indicates the sign of the user polarity score, 
as defined by \citet{barbera2015birds} and described in Section~\ref{sec:measures} 
(grey$\,=\,$negative$\,=\,$democrat, yellow$\,=\,$positive$\,=\,$republican).
The difference between the two groups of datasets is stark: 
production and consumption polarities are highly correlated for \controversial datasets, which means that users indeed tend to consume content with political leaning aligned to their own.
The same does not hold for the \noncontroversial group, where the correlations are low to non-existent.

How do the production and consumption polarities align with user polarity scores?
To explore this, let us turn to the one-di\-men\-sio\-nal subplots that accompany each scatter-plot.
The subplot along the $x$-axis ($y$-axis) shows the distributions for production (consumption) polarity for democrats and republicans --- 
as before, defined in terms of the sign of user polarity~\cite{barbera2015birds}.
We observe that the production and consumption polarities for the \controversial datasets 
exhibit clearly separated and bi-modal distributions, 
while the distributions very much coincide for the \noncontroversial datasets. 
This kind of bimodal distribution is also indicative of a divide in the leaning of the content produced and consumed.

Furthermore, let us note that, when the distributions of production and consumption polarities are compared with the source polarity scores in Figure~\ref{fig:content_polarity_distribution}, they appear quite different.
The production/consumption polarities are more concentrated towards the middle of the spectrum (i.e., there are few very extreme users), and the modes themselves are relatively far from the extremes.
In addition, the concentration of the distributions show a preference for one leaning when compared to the distribution of source polarities.
This preference can be attributed to personal choice of the user (for the production), and also to network effects such as homophily and network correlation (for the consumption).

\begin{figure*}[t]
\begin{minipage}{.19\linewidth}
\centering
\subfloat[]{\label{}\includegraphics[width=\textwidth]{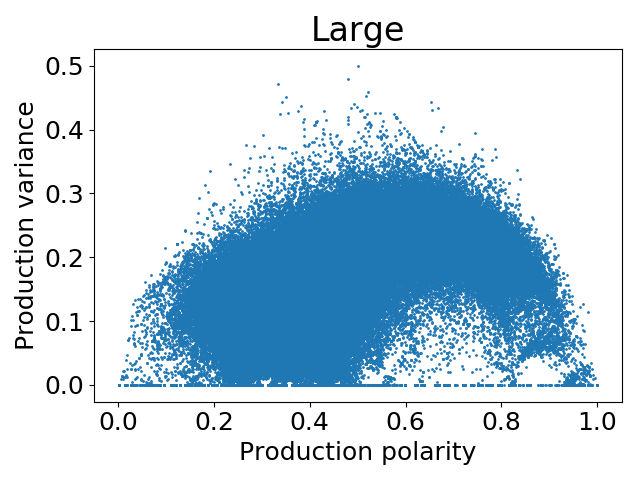}}
\end{minipage}%
\begin{minipage}{.19\linewidth}
\centering
\subfloat[]{\label{}\includegraphics[width=\textwidth]{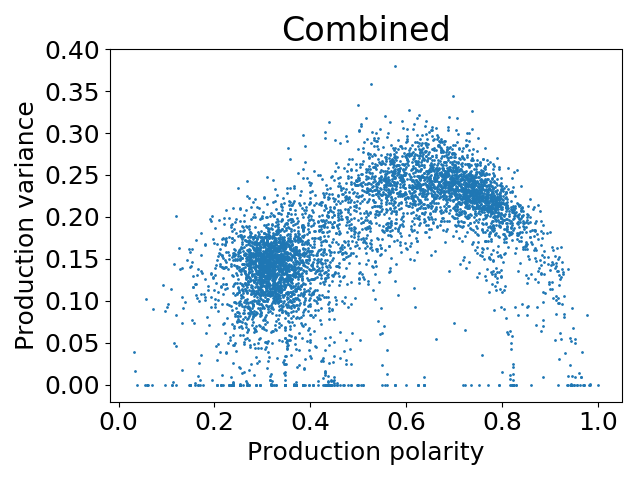}}
\end{minipage}
\begin{minipage}{.19\linewidth}
\centering
\subfloat[]{\label{}\includegraphics[width=\textwidth]{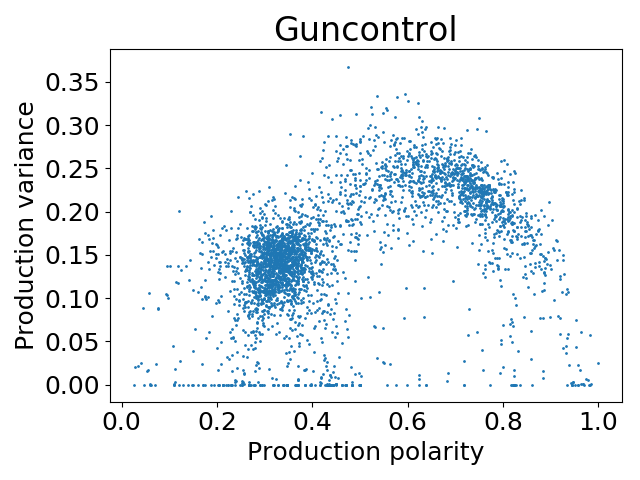}}
\end{minipage}
\begin{minipage}{.19\linewidth}
\centering
\subfloat[]{\label{}\includegraphics[width=\textwidth]{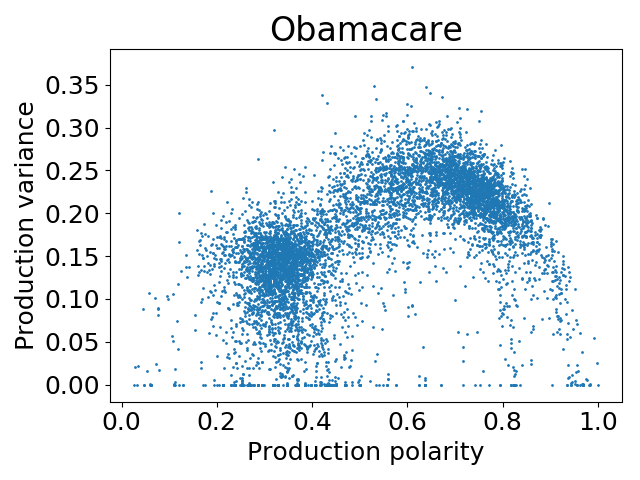}}
\end{minipage}
\begin{minipage}{.19\linewidth}
\centering
\subfloat[]{\label{}\includegraphics[width=\textwidth]{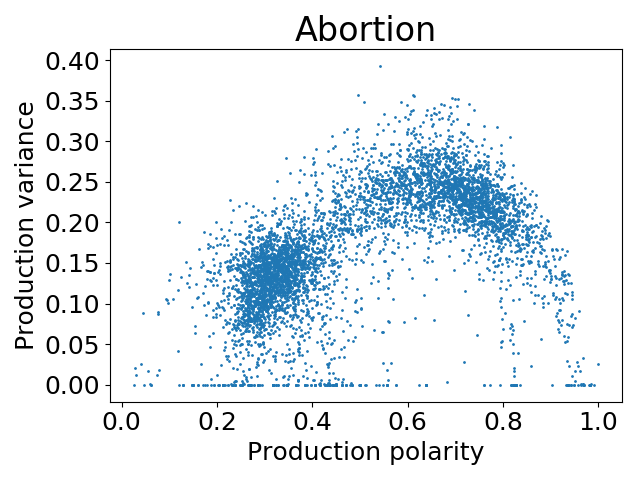}}
\end{minipage}%
\par\medskip
\begin{minipage}{.19\linewidth}
\centering
\subfloat[]{\label{}\includegraphics[width=\textwidth]{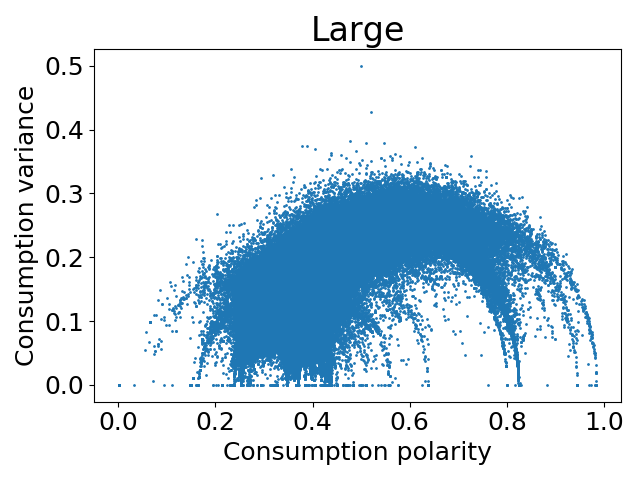}}
\end{minipage}
\begin{minipage}{.19\linewidth}
\centering
\subfloat[]{\label{}\includegraphics[width=\textwidth]{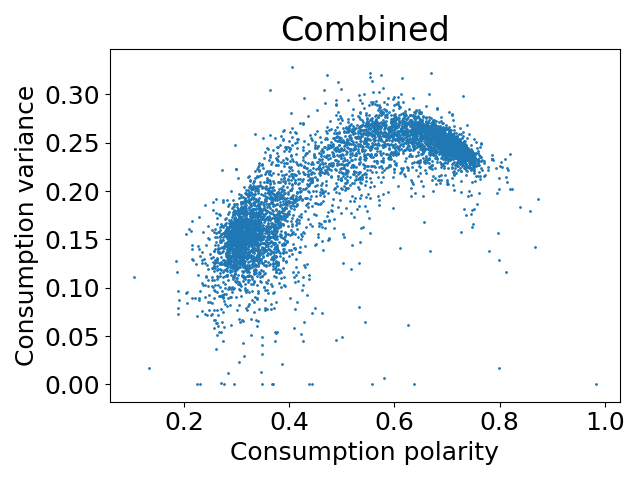}}
\end{minipage}
\begin{minipage}{.19\linewidth}
\centering
\subfloat[]{\label{}\includegraphics[width=\textwidth]{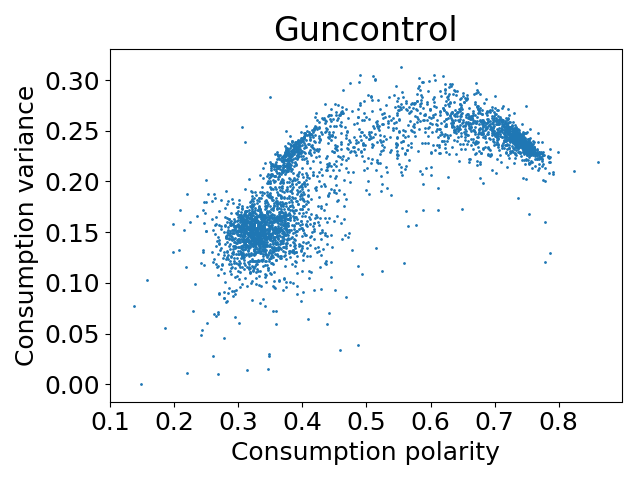}}
\end{minipage}
\begin{minipage}{.19\linewidth}
\centering
\subfloat[]{\label{}\includegraphics[width=\textwidth]{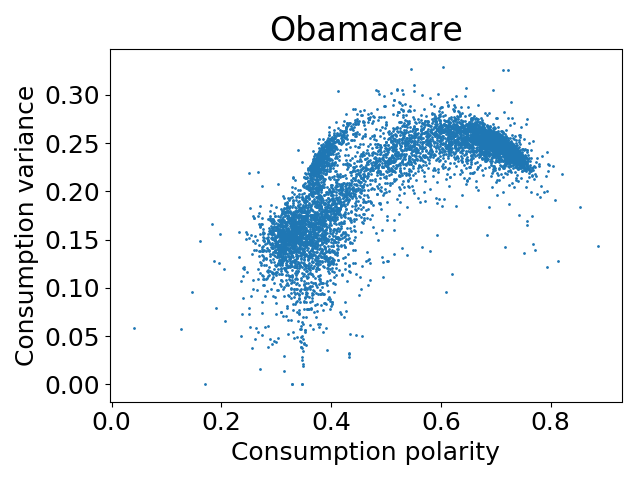}}
\end{minipage}
\begin{minipage}{.19\linewidth}
\centering
\subfloat[]{\label{}\includegraphics[width=\textwidth]{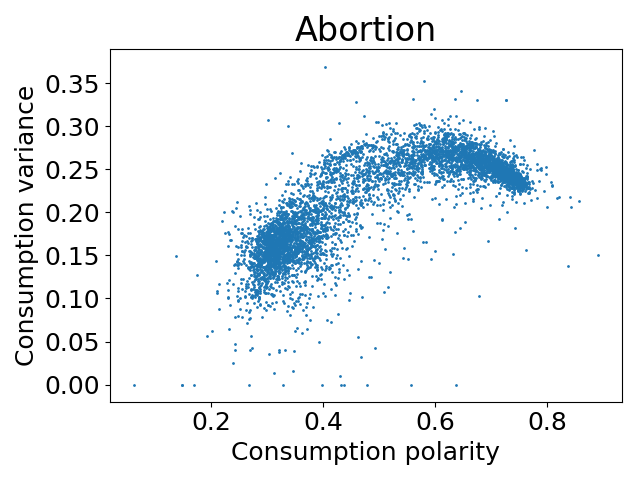}}
\end{minipage}
\par\medskip
\caption{Top: Production polarity variance vs. production polarity (mean).
Bottom: Consumption polarity variance vs. consumption polarity (mean).}
\label{fig:avg_var}
\vspace{-\baselineskip}
\end{figure*}

Finally, we examine the variance of the production and consumption polarities.
We ask whether users who are more partisan also present a \emph{lower} variance in their polarities, which means they produce and consume content from a narrower spectrum of sources.
Figure~\ref{fig:avg_var} shows the consumption and production variance of each user ($y$-axis) against the respective (mean) polarity measure. 
The plot shows a clear ``downward U'' trend, which confirms the aforementioned hypothesis: bipartisan users follow news sources with a wider spread of political leaning, rather than just picking from the center, which makes their news diet qualitatively different from partisan users.
We obtain similar results when we examine the variance of production and consumption polarities as a function of user polarity score~\citep{barbera2015birds} (omitted due to space constraints).
The consistency of these results reinforces the validity of our production and consumption polarity metrics.

\vspace{-0.2cm}
\subsection{Analysis of partisan users}
\label{subsec:partisans}

Recall that a \deltapartisan user is one who tends to produce content exclusively from one side of the political spectrum.
In this section, we study how partisan users differ from bipartisan users.
We focus on three main elements for the comparison:
\begin{squishlist}
\item[(a)] {\em Network}: PageRank (global measure of centrality), clustering coefficient (local measure of centrality), and absolute user polarity (higher values indicate higher polarization).
\item[(b)] {\em Profile}: number of followers (proxy for popularity), number of friends, number of tweets (proxy for activity), age on Twitter (number of weeks the user has been on Twitter).
\item[(c)] {\em Interaction}: retweet/favorite rate, retweet/favorite volume.
\end{squishlist}

Partisans and bipartisans are parameterized by a threshold~$\delta$, 
and we consider different values for $\delta$ between $0.20$ and $0.45$ in steps of $0.05$.
For each value of $\delta$, we explore the value distribution of the above features for the two groups of users and test whether they are different.
Table~\ref{tab:comparison_partisans_gatekeepers} (second column) summarizes the results for partisan users and lists the features for which the difference is statistically significant on a majority of the datasets.
A ``\cmark'' in the table means that the property (e.g., PageRank) is significantly higher for partisans
 for at least 4 of the 6 values of the $\delta$ threshold,
for most of the datasets (In most cases we find consistent behavior across all datasets).\footnote{Significance tested using Welch's $t$-test for equality of means ($p < 0.001$)~\cite{welch1947generalization}.}
A ``\cmark (-)'' means that the property is significantly lower for partisans.
A ``\xmark'' indicates we find no significant difference.

\begin{figure*}[t]
\begin{minipage}{.19\linewidth}
\centering
\subfloat[]{\label{}
\includegraphics[width=\textwidth, height=\textwidth]{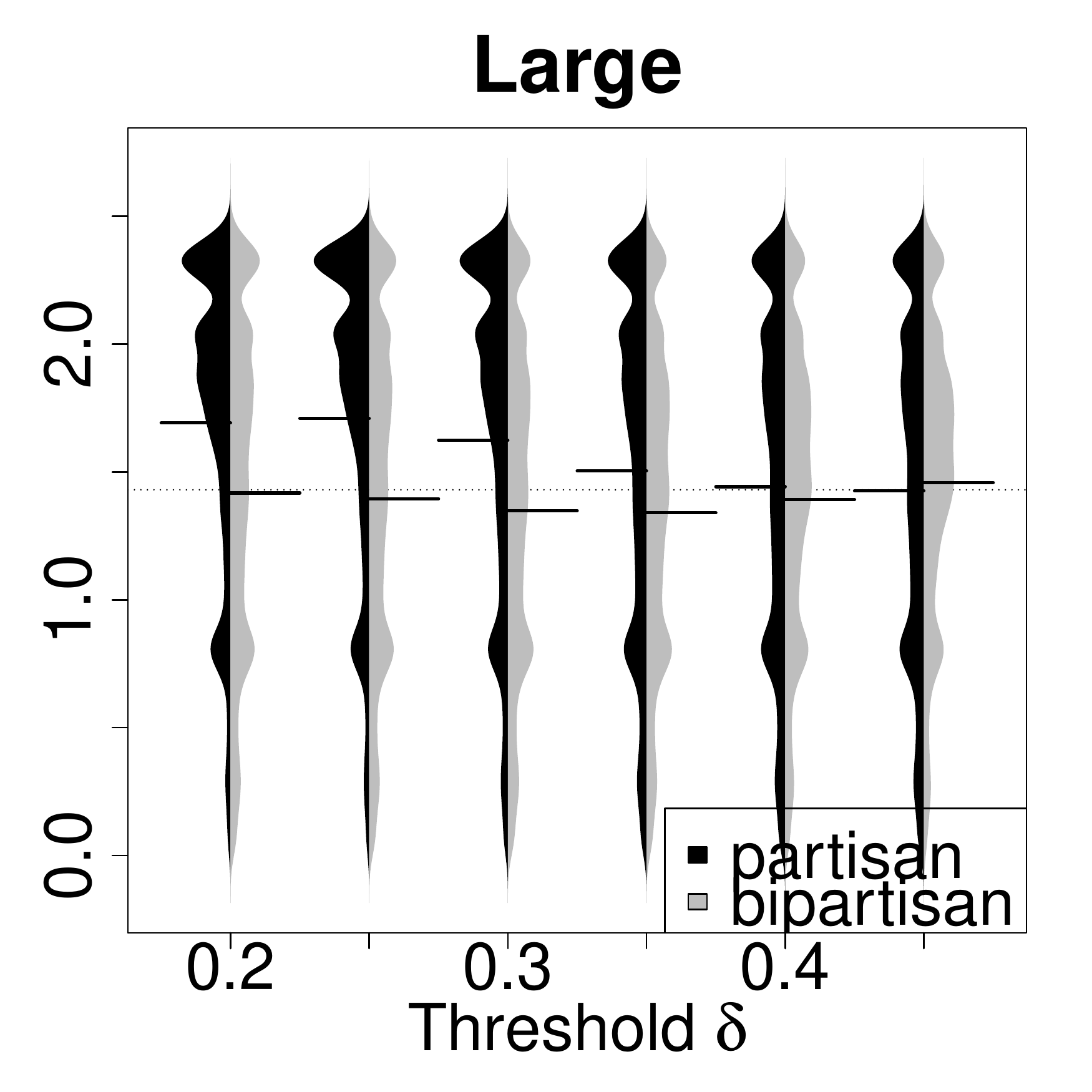}}
\end{minipage}%
\begin{minipage}{.19\linewidth}
\centering
\subfloat[]{\label{}
\includegraphics[width=\textwidth, height=\textwidth]{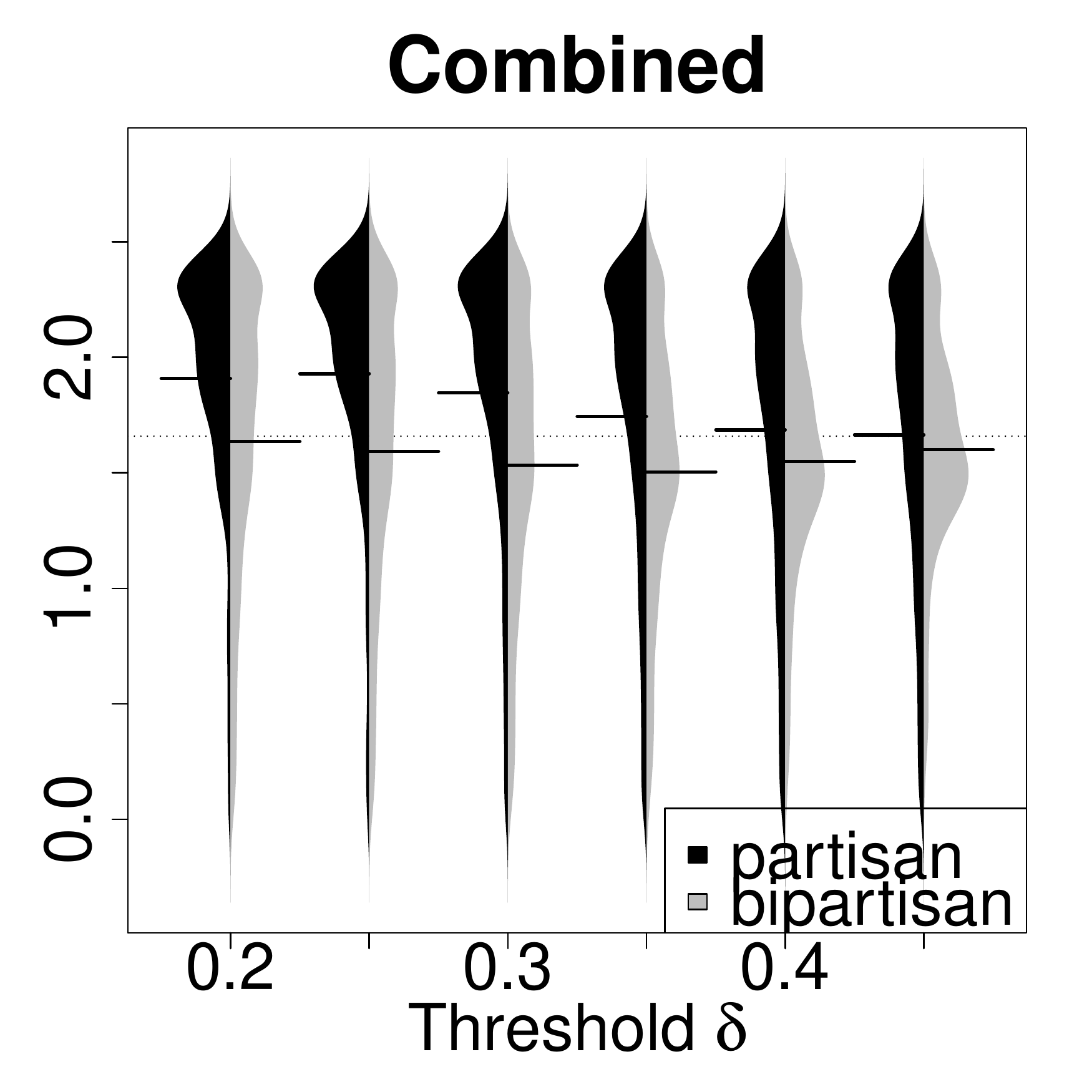}}
\end{minipage}%
\begin{minipage}{.19\linewidth}
\centering
\subfloat[]{\label{}
\includegraphics[width=\textwidth, height=\textwidth]{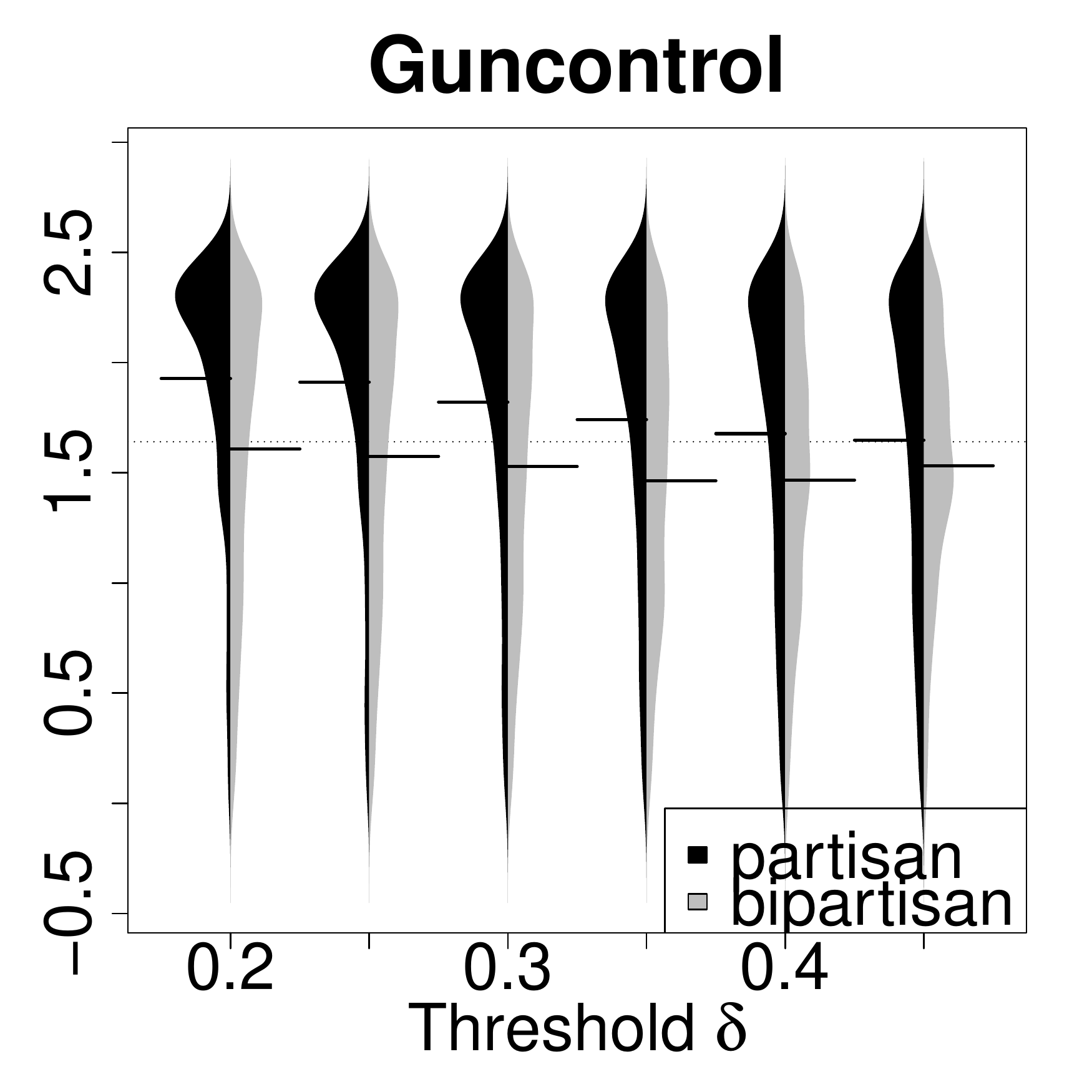}}
\end{minipage}
\begin{minipage}{.19\linewidth}
\centering
\subfloat[]{\label{}
\includegraphics[width=\textwidth, height=\textwidth]{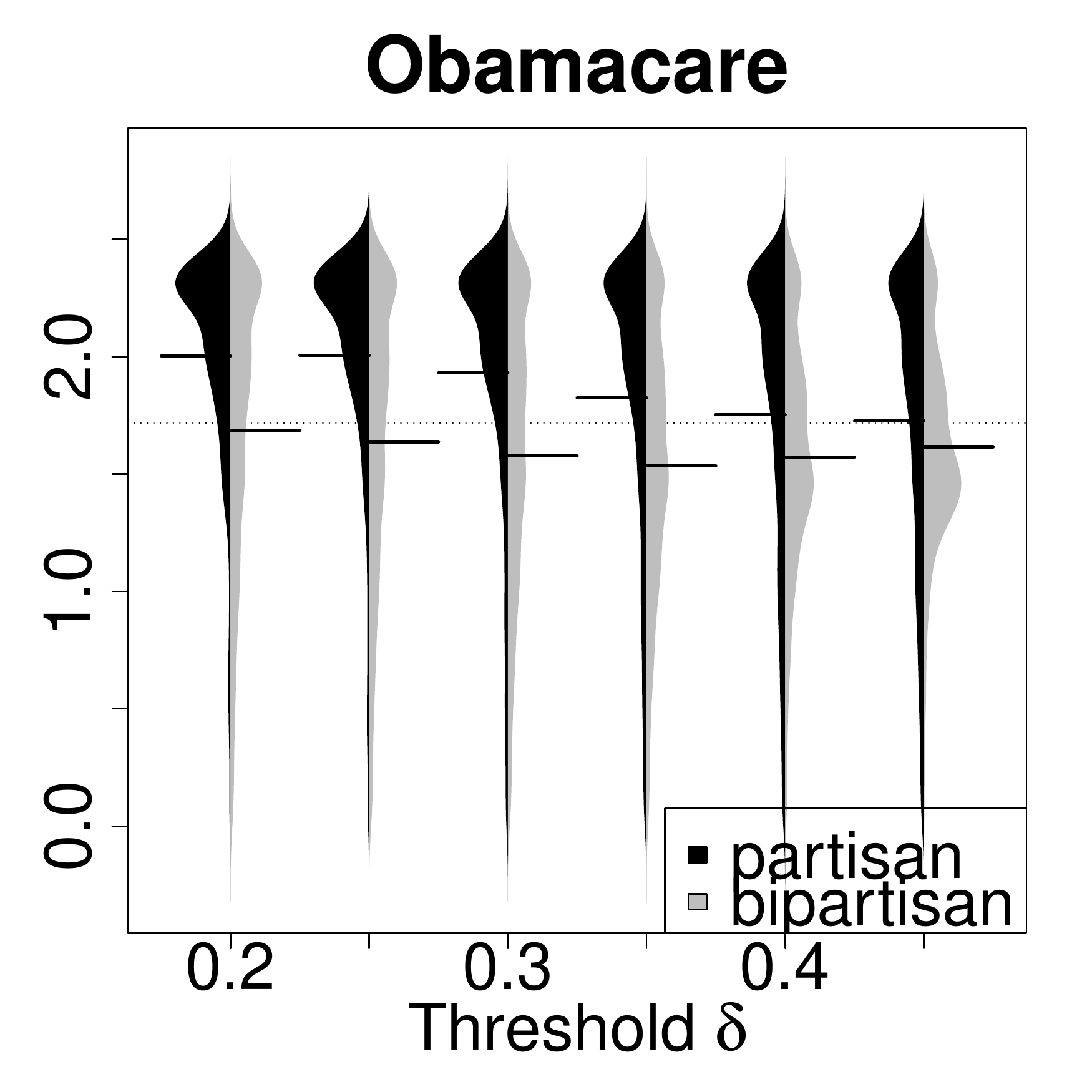}}
\end{minipage}
\begin{minipage}{.19\linewidth}
\centering
\subfloat[]{\label{}
\includegraphics[width=\textwidth, height=\textwidth]{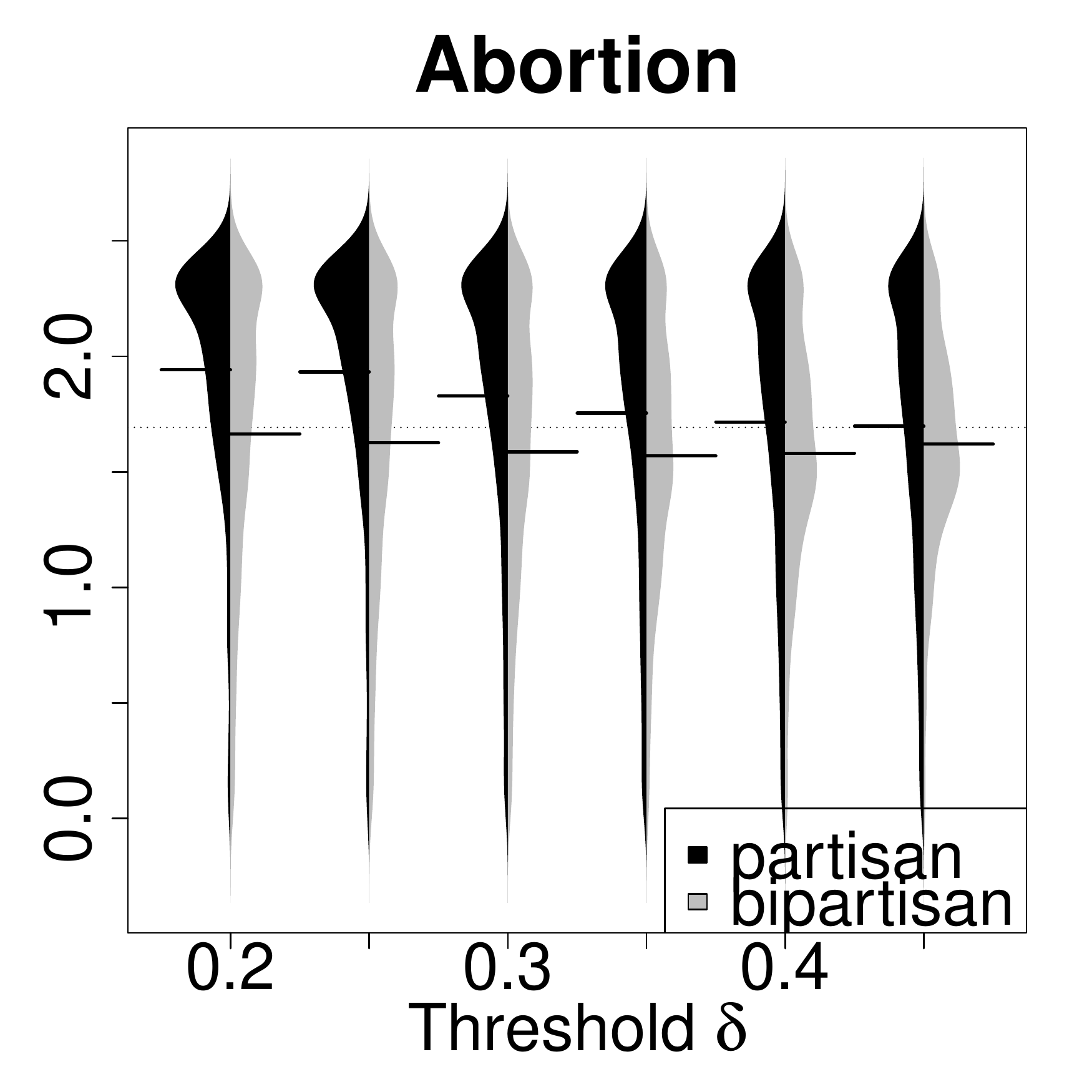}}
\end{minipage}\par\medskip
\caption{Absolute value of the user polarity scores for \deltapartisan and \deltabipartisan users.}
\label{fig:polarity}
\vspace{-\baselineskip}
\end{figure*}

\begin{figure*}[t]
\begin{minipage}{.19\linewidth}
\centering
\subfloat[]{\label{}
\includegraphics[width=\textwidth, height=\textwidth]{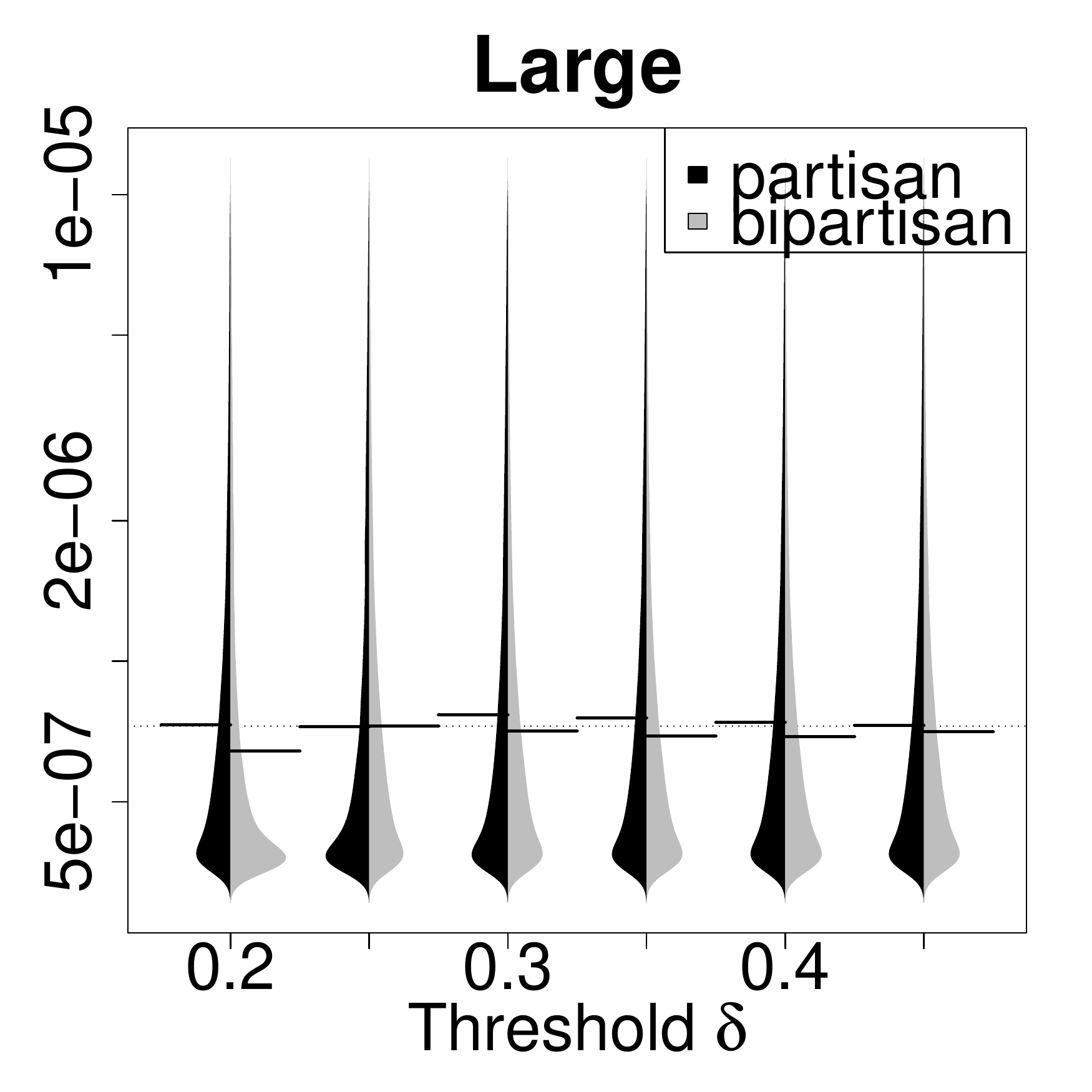}}
\end{minipage}%
\begin{minipage}{.19\linewidth}
\centering
\subfloat[]{\label{}
\includegraphics[width=\textwidth, height=\textwidth]{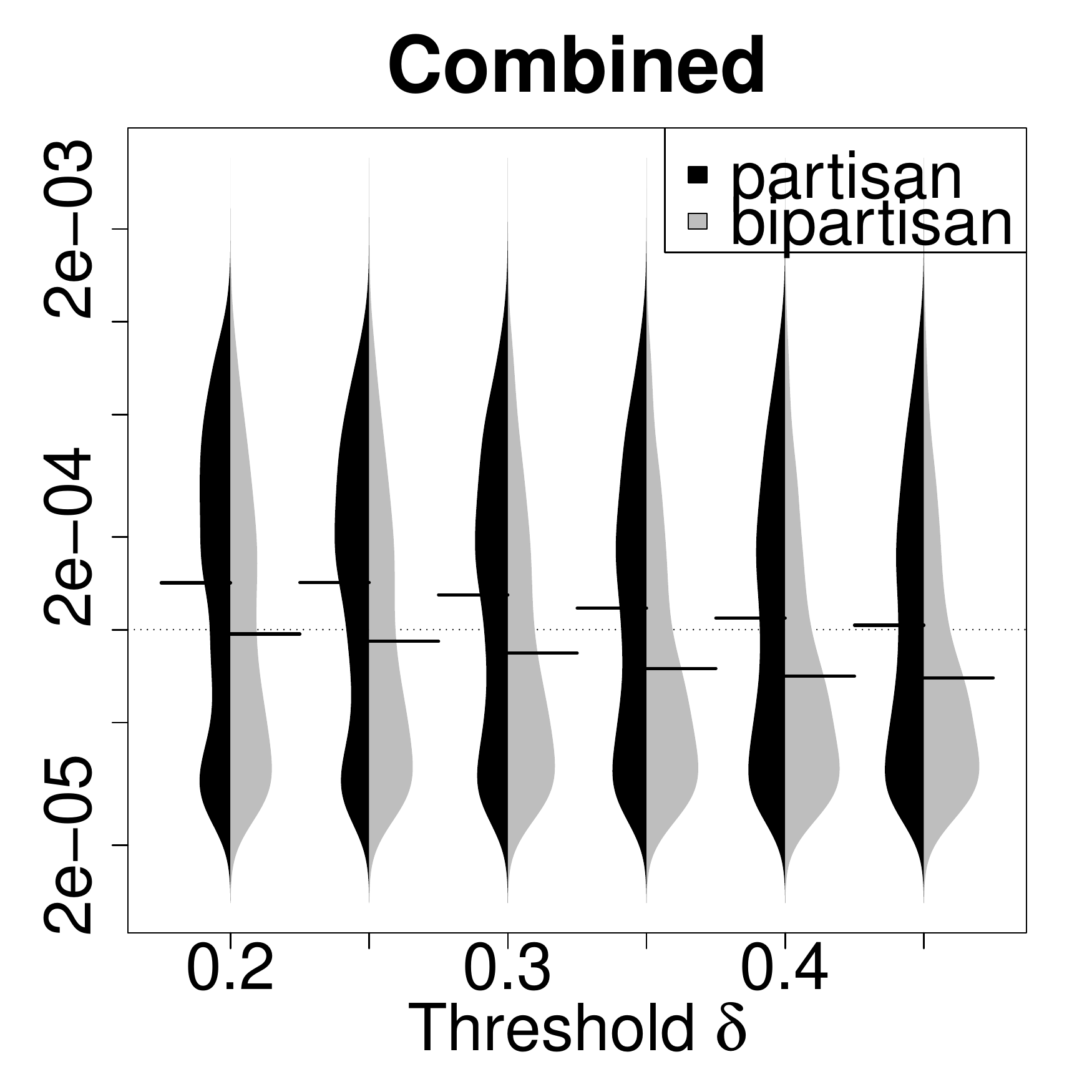}}
\end{minipage}%
\begin{minipage}{.19\linewidth}
\centering
\subfloat[]{\label{}
\includegraphics[width=\textwidth, height=\textwidth]{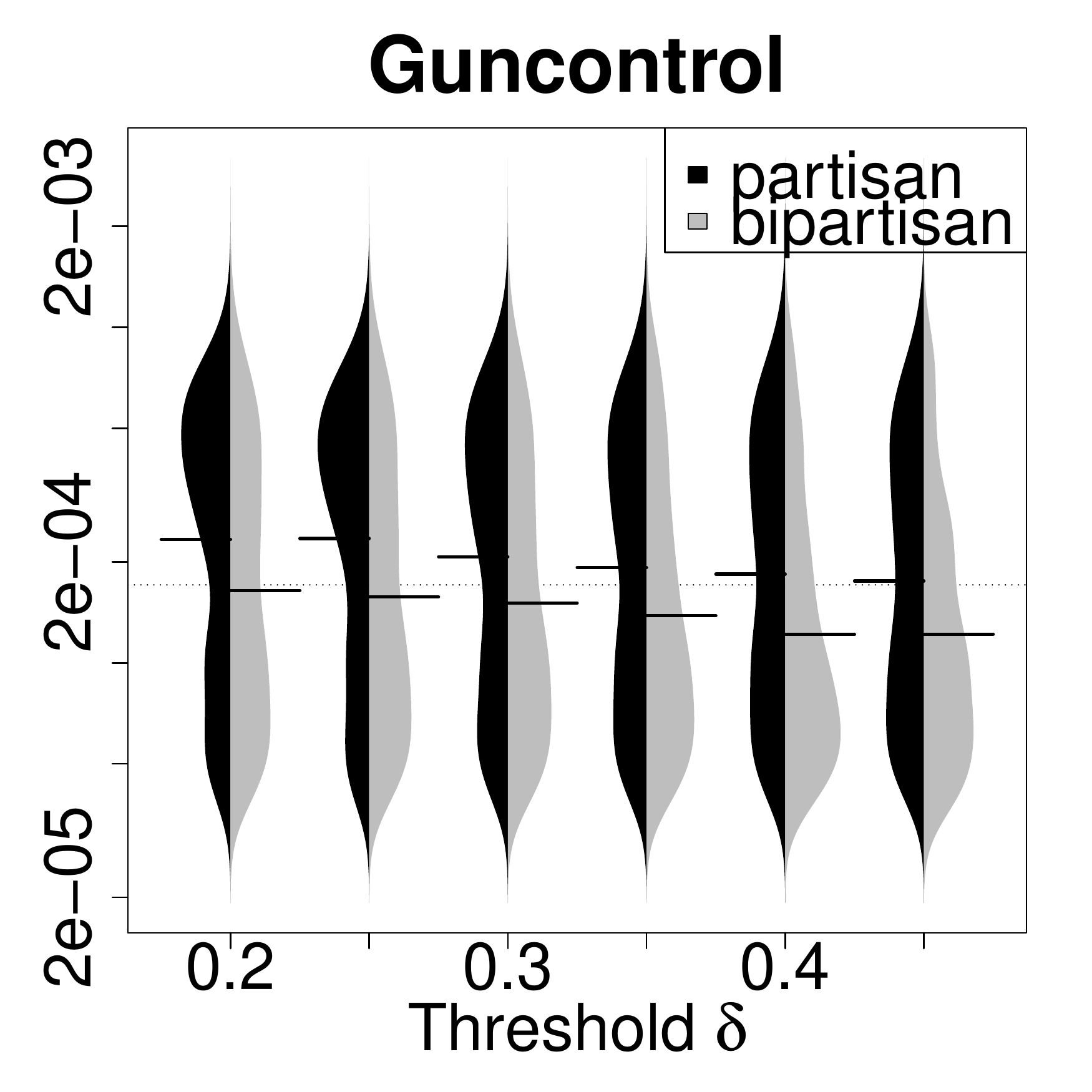}}
\end{minipage}
\begin{minipage}{.19\linewidth}
\centering
\subfloat[]{\label{}
\includegraphics[width=\textwidth, height=\textwidth]{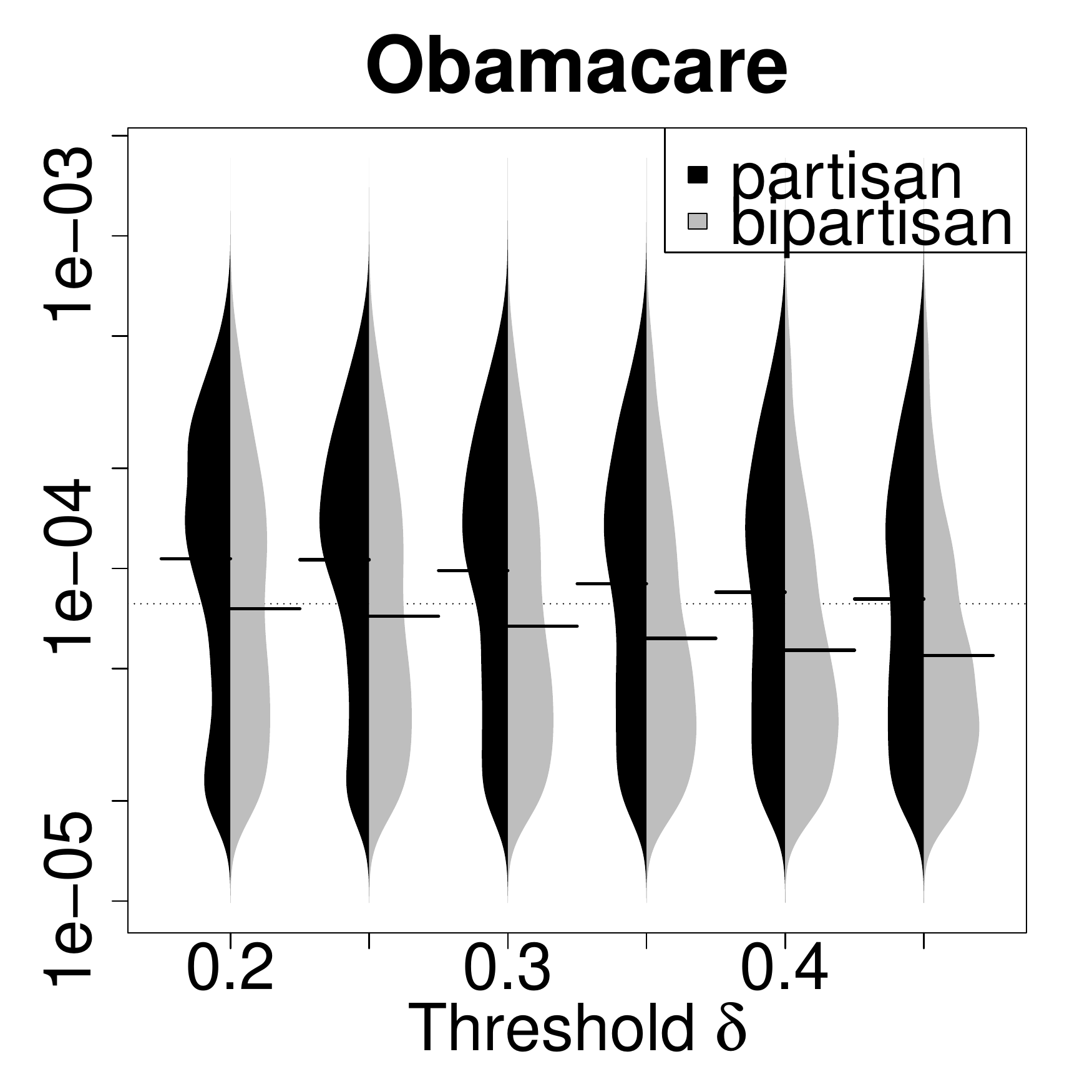}}
\end{minipage}
\begin{minipage}{.19\linewidth}
\centering
\subfloat[]{\label{}
\includegraphics[width=\textwidth, height=\textwidth]{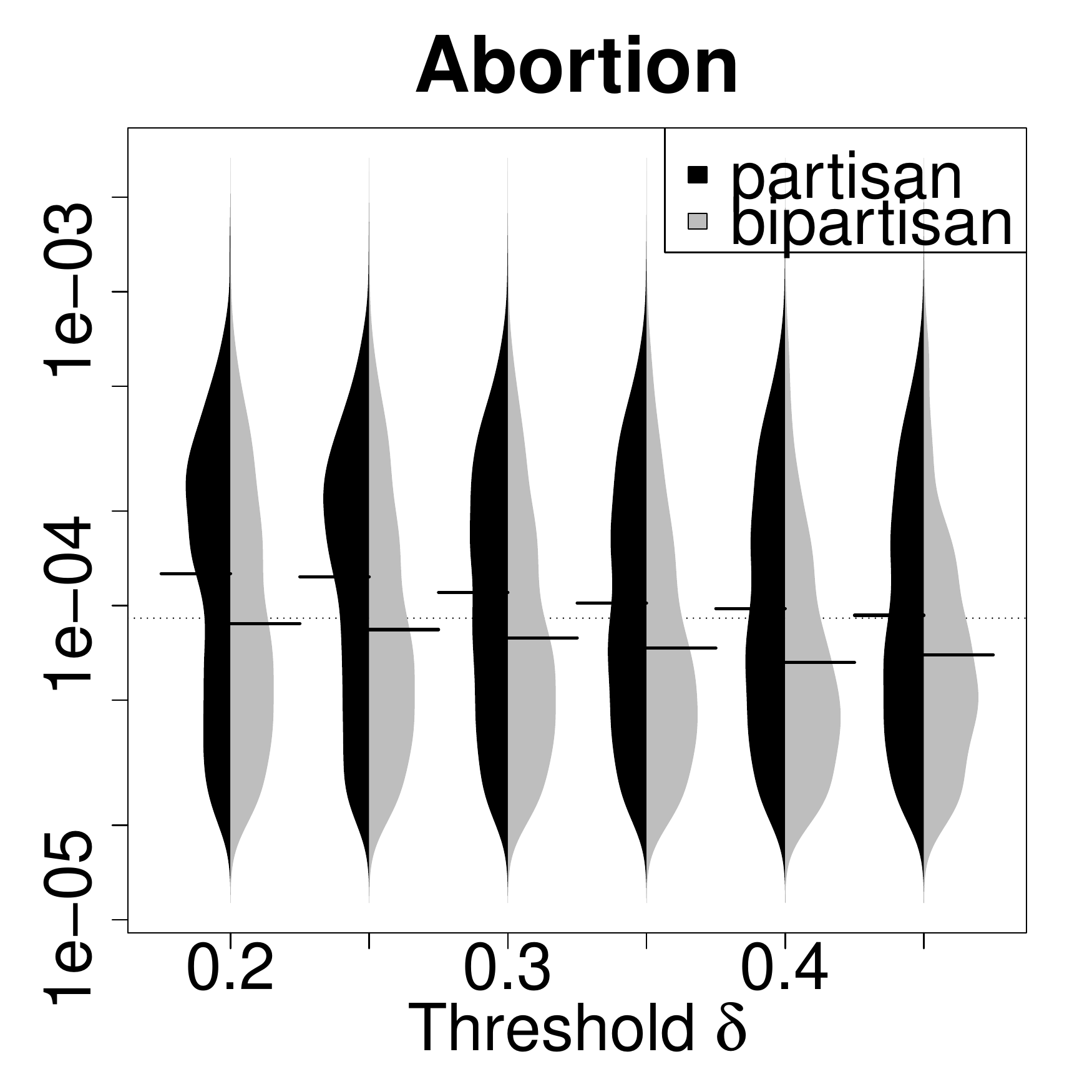}}
\end{minipage}\par\medskip
\caption{Pagerank for \deltapartisan and \deltabipartisan users.}
\label{fig:pagerank1}
\vspace{-\baselineskip}
\end{figure*}

\begin{figure*}[t]
\begin{minipage}{.19\linewidth}
\centering
\subfloat[]{\label{}
\includegraphics[width=\textwidth, height=\textwidth]{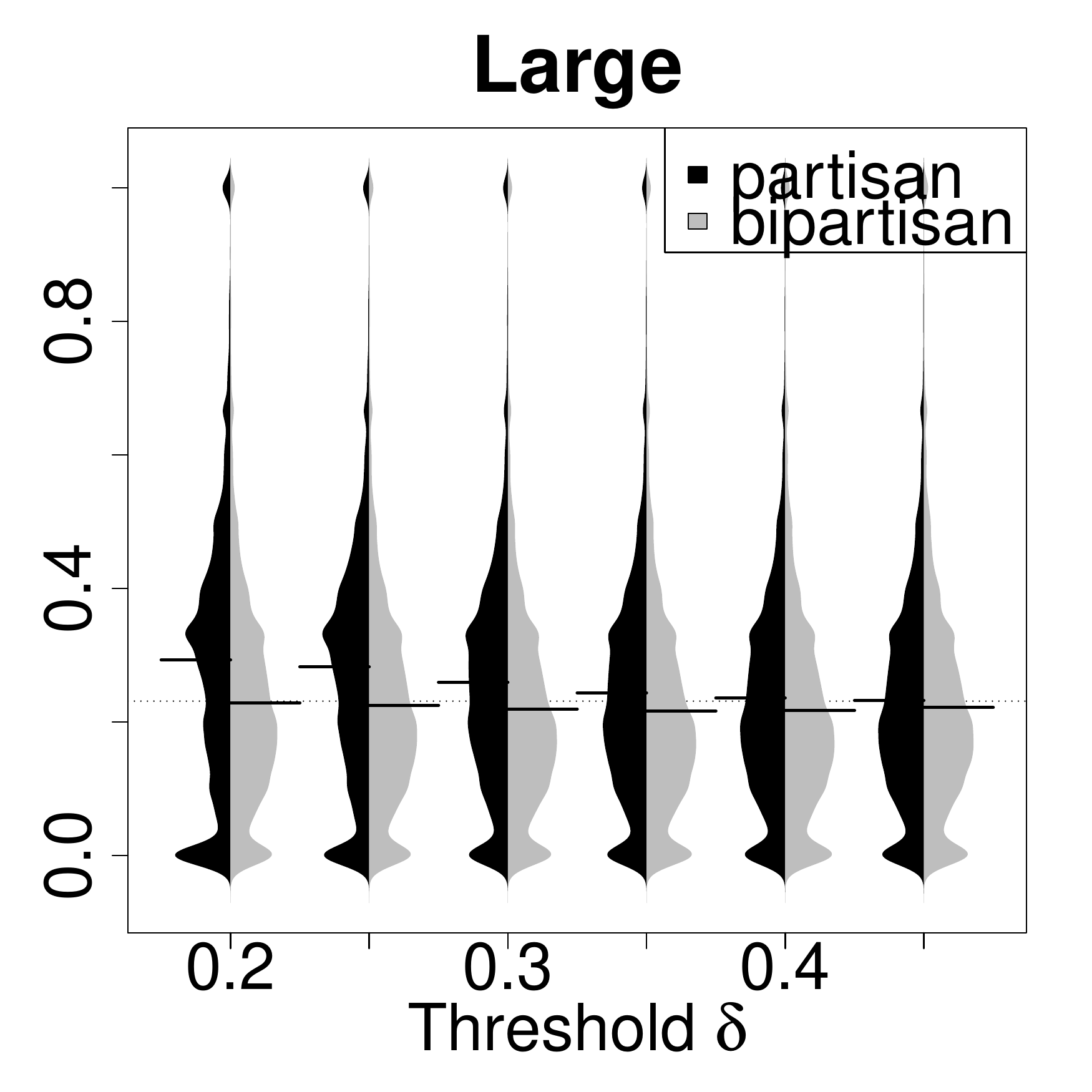}}
\end{minipage}%
\begin{minipage}{.19\linewidth}
\centering
\subfloat[]{\label{}
\includegraphics[width=\textwidth, height=\textwidth]{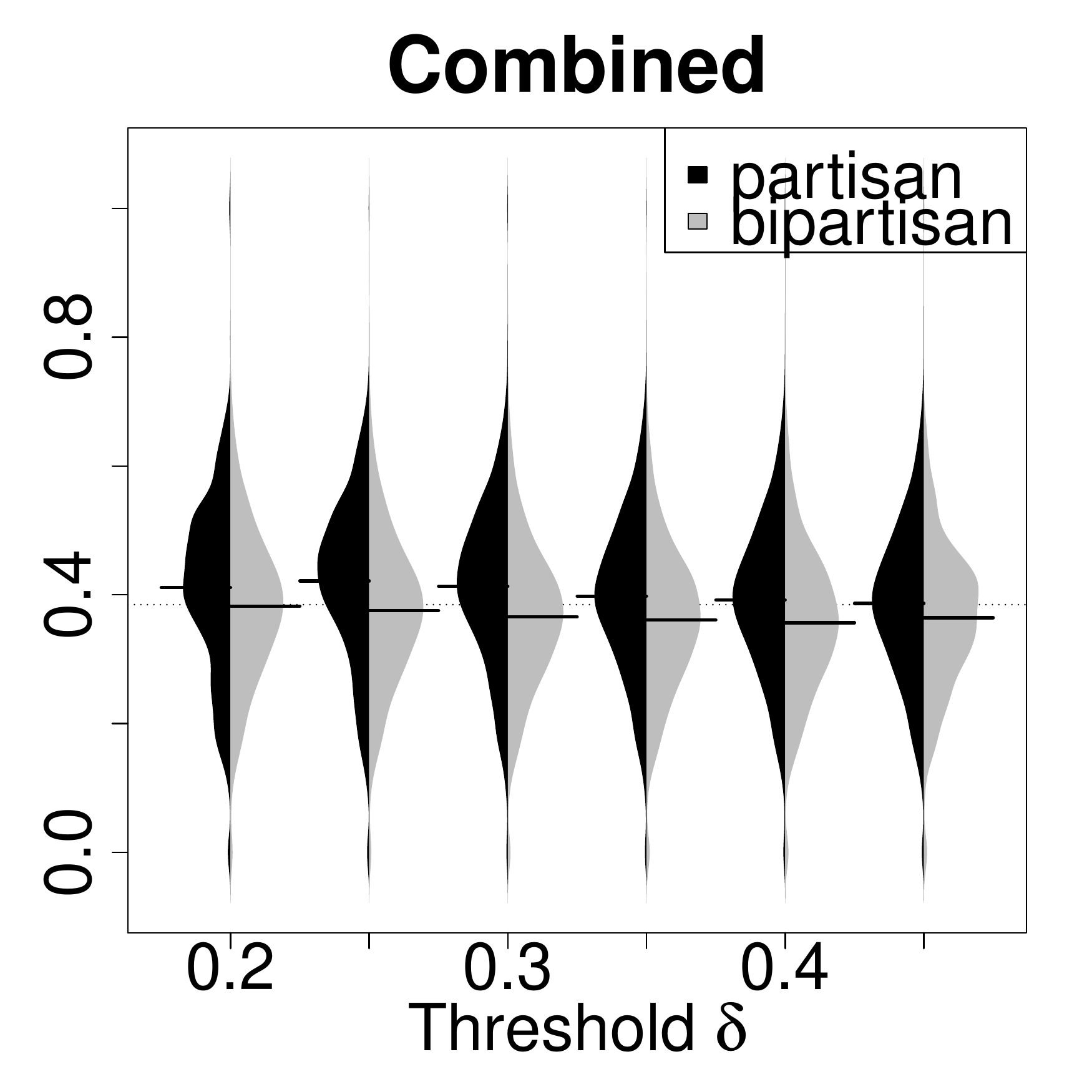}}
\end{minipage}%
\begin{minipage}{.19\linewidth}
\centering
\subfloat[]{\label{}
\includegraphics[width=\textwidth, height=\textwidth]{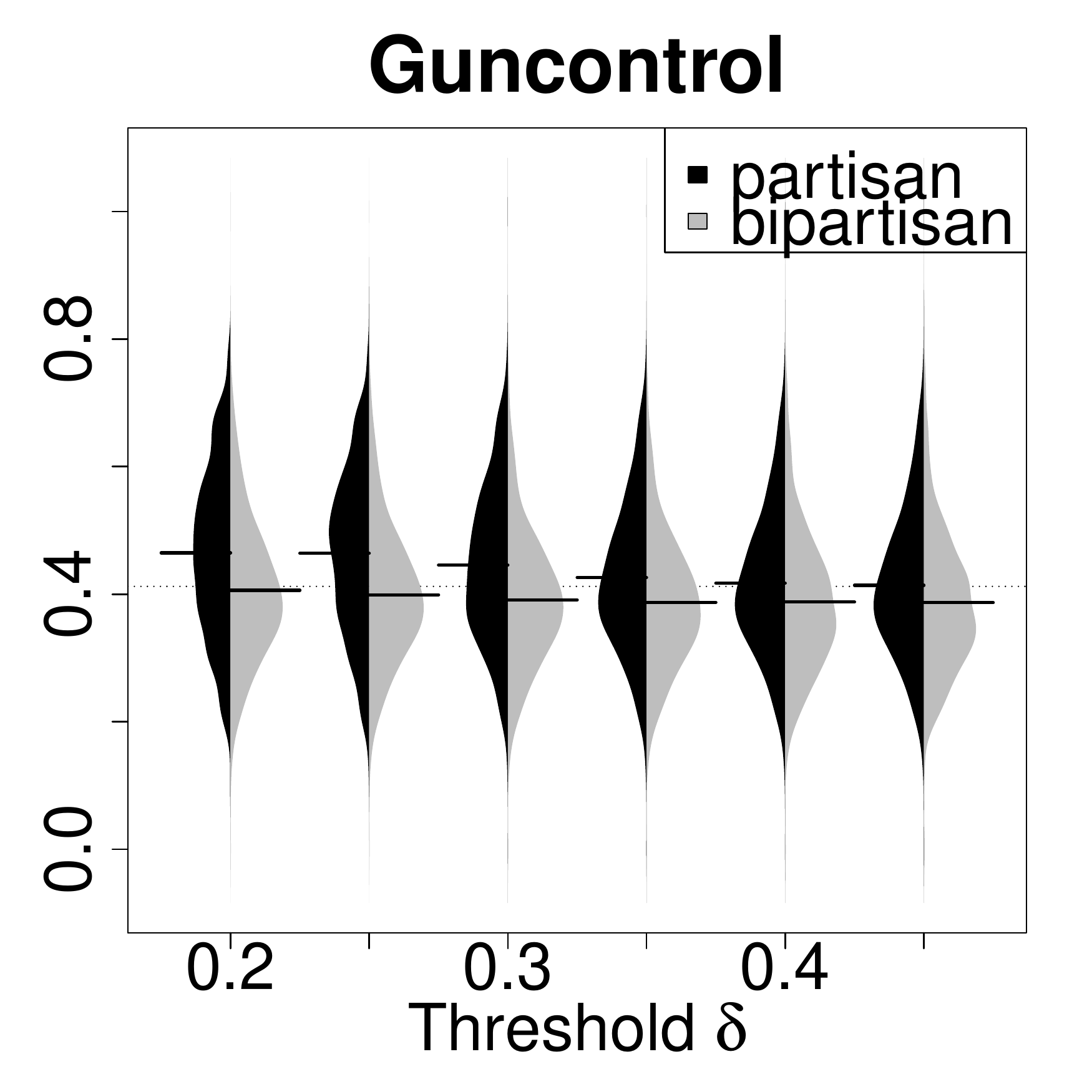}}
\end{minipage}
\begin{minipage}{.19\linewidth}
\centering
\subfloat[]{\label{}
\includegraphics[width=\textwidth, height=\textwidth]{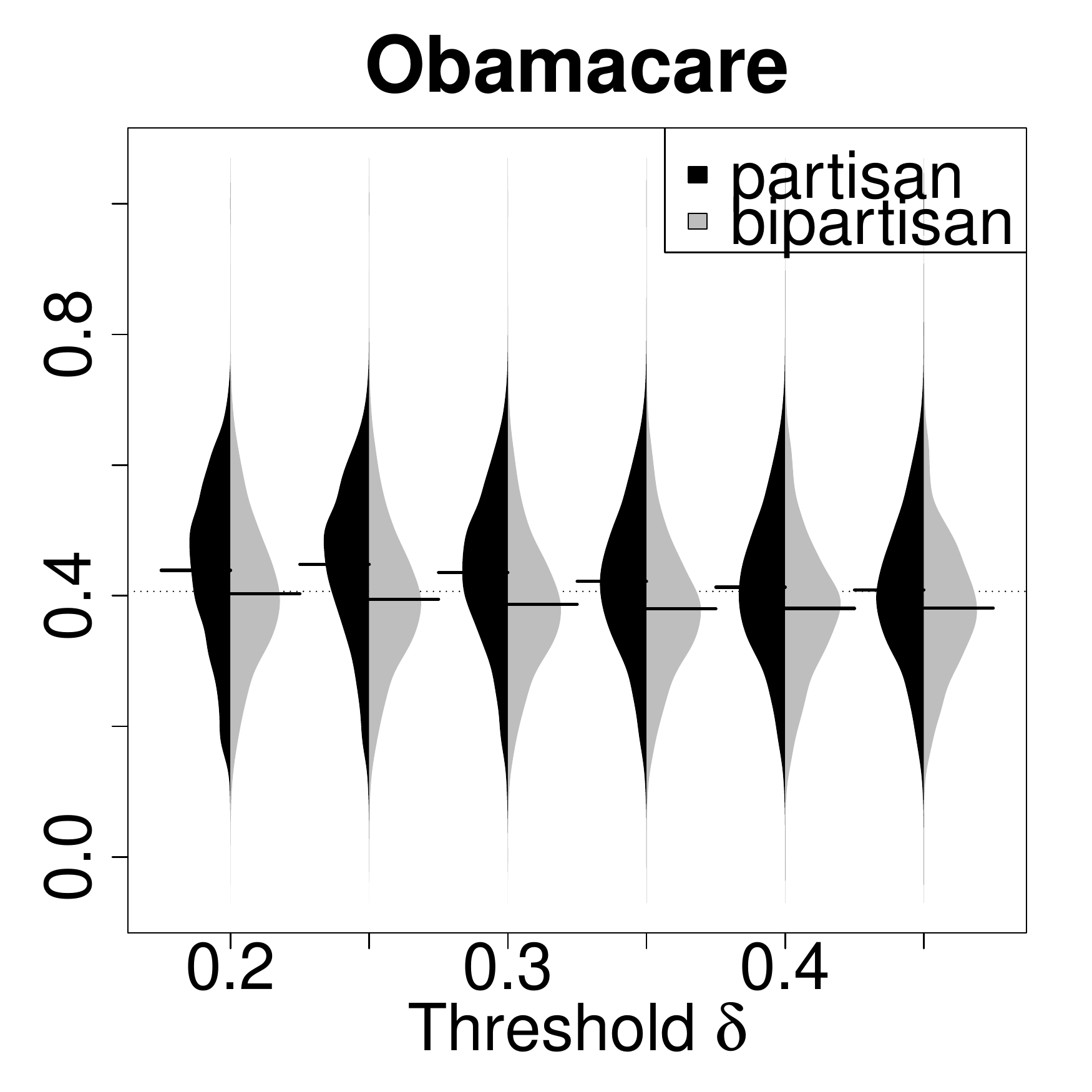}}
\end{minipage}
\begin{minipage}{.19\linewidth}
\centering
\subfloat[]{\label{}
\includegraphics[width=\textwidth, height=\textwidth]{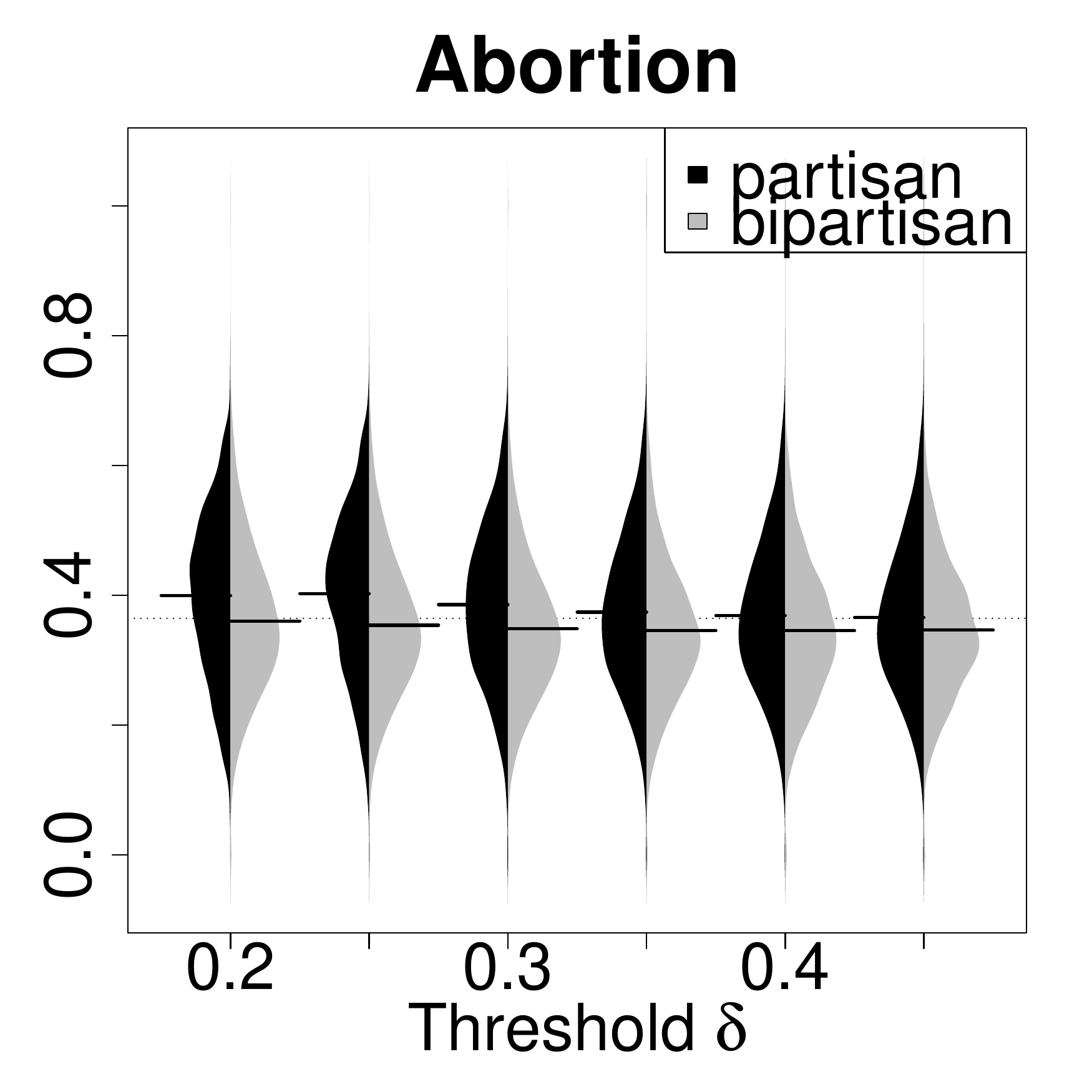}}
\end{minipage}\par\medskip
\caption{Clustering Coefficient for \deltapartisan and \deltabipartisan users.}
\label{fig:clustering_coefficient}
\vspace{-\baselineskip}
\end{figure*}

For some of the features that exhibit significantly different distributions between the two groups, the distributions are shown in 
Fig.~\ref{fig:polarity} (user polarity),
Fig.~\ref{fig:pagerank1} (PageRank).
and Fig.~\ref{fig:clustering_coefficient} (clustering coefficient).
Each figure shows a set of beanplots,\footnote{A beanplot is an alternative to the boxplot for visual comparison of univariate data among groups.} one for each \controversial dataset.
Each beanplot shows the estimated probability density function for a measure computed on the dataset, the individual observations are shown as small white lines in a one-dimensional scatter plot, and the mean as a longer black line.
The beanplot is divided into two groups, one for partisan users (left/dark) and one for bi-partisan ones (right/light).


Considering absolute user polarity scores, 
partisan users are significantly more polarized than bipartisan ones, as shown in Figure~\ref{fig:polarity}.
We see that partisan users enjoy a more central position in the network, 
indicated by higher PageRank (Figure~\ref{fig:pagerank1}).
Similarly, partisan users are more connected to their own community, 
indicated by a higher clustering coefficient (Figure~\ref{fig:clustering_coefficient}).
Finally, their tweets are more appreciated, i.e., 
a higher fraction of their tweets receives a retweet, albeit the effect size is smaller in this case (figure not shown).
Similar trends hold for the number of retweets and the number of favorites (omitted due to space constraints).
These results are consistent irrespective of the value of the $\delta$ threshold used to define \deltapartisan users.
We do not find any consistent trend across datasets in terms of profile features. 
Table~\ref{tab:comparison_partisans_gatekeepers} shows a summary of these trends.



\begin{table}[]
\centering
\caption{Comparison of various features for partisans \& bipartisans and gatekeepers \& non-gatekeepers. A \cmark indicates that the corresponding feature is significantly higher for the group of the column ($p < 0.001$) for at least 4 of the 6 thresholds $\delta$ used, for most datasets. A minus next to the checkmark (-) indicates that the feature is significantly lower.}
\label{tab:comparison_partisans_gatekeepers}
\begin{small}
\begin{tabular}{l c c}
	\toprule
         Features                                                        & Partisans & Gatekeepers \\
	\midrule
PageRank                                                         &   \cmark       &    \cmark         \\
clustering coefficient                                           &   \cmark       &  \txtwhite{(-)} \cmark (-) \\
user polarity                                                    &   \cmark       &   \txtwhite{(-)} \cmark (-) \\
degree                                                           &   \cmark       &    \cmark         \\
retweet rate                                                     &   \cmark       &    \xmark         \\
retweet volume                                                   &   \cmark       &    \xmark         \\
favorite rate                                                    &   \cmark       &    \xmark         \\
favorite volume                                                  &   \cmark       &    \xmark         \\
\# followers                                                     &  \xmark        &    \xmark         \\
\# friends                                                       &  \xmark        &    \xmark         \\
\# tweets                                                        &  \xmark        &    \xmark         \\
age on Twitter                                                   &  \xmark        &    \xmark        \\
\bottomrule 
\end{tabular}
\end{small}
\vspace{-\baselineskip}
\end{table}

\subsection{Gatekeepers of information}
\label{subsec:gatekeepers}

We now turn our attention to \deltakeeper users, 
i.e., users who consume more central content than they produce.
As in the previous section, we vary $\delta$ between $0.20$ and $0.45$ in intervals of~$0.05$ and compare
gatekeepers with other users who are not gatekeepers.
Due to space constraints, 
we do not show beanplots for the gatekeepers.
We only show a summary of results in Table~\ref{tab:gatekeepers}.

Gatekeepers, like partisans, occupy positions with high centrality in the network, 
i.e., higher than average PageRank and in-degree.
However, differently from the rest of the side they align with, 
they show a lower clustering coefficient, 
an indication that they are not completely embedded in a single community.
Given that they receive content also from the opposing side, this result is to be expected: 
most of the links that span the two communities will remain open (i.e., not form a triangle).
Similarly, their polarity score is on average less extreme than the rest of their group.

Differently from the partisans, 
we could not find consistent trends for interaction features
such as retweet and favorite rate and volume.
Profile features are also not consistently different for gatekeepers.
The results are reported in Table~\ref{tab:comparison_partisans_gatekeepers}.

Finally, given that both partisans and gatekeepers sport higher centrality, we compare their PageRank values directly
and find that there is a significant difference: partisans have a higher PageRank compared to gatekeepers (figure not shown).
This effect is more pronounced for higher values of the threshold $\delta$, possibly suggesting that, even among users who produce polarized content, purity  (not following users of the opposite side) is rewarded.

\begin{table}[]
\centering
\caption{Comparison between \deltakeeper users and a random sample of normal users. A \cmark indicates that the corresponding property is significantly higher for gatekeepers ($p < 0.001$) for at least 4 of the 6 thresholds $\delta$ used. A minus next to the checkmark (-) indicates that the property is significantly lower.}
\label{tab:gatekeepers}
\begin{small}
\begin{tabular}{l c c c c}
\toprule
           & PageRank & Degree & CC  & Polarity \\
\midrule
\guncontrol & \cmark      & \cmark    & \cmark (-) &  \cmark (-)  \\
\obamacare  & \cmark      &      & \cmark (-) &  \cmark (-) \\
\election     & \cmark      & \cmark    & \cmark (-) &  \cmark (-) \\
\abortion   & \cmark      & \cmark    & \cmark (-) &   \cmark (-) \\
\largetw      & \cmark      & \cmark    & \cmark (-) &  \cmark (-) \\
\bottomrule
\end{tabular}
\end{small}
\vspace{-\baselineskip}
\end{table}

\subsection{Prediction}
\label{section:prediction}

Given that partisans and gatekeepers present markedly different characteristics in terms of network and content, can we predict a user's role as partisan and gatekeeper without knowledge of their production and consumption polarities?
That is, how evident is their role in the discussion just by examining their network, and profile features?
We train a Random Forest classifier on the \controversial datasets, and use the following features for each user:
\begin{squishlist}
\item[$-$] \emph{Network features}: PageRank, degree, clustering coefficient;
\item[$-$] \emph{Profile features}: number of tweets, of followers, of friends, age on Twitter;
\item[$-$] \emph{Tweet features}: $n$-grams with tf-idf weights from their tweets.
\end{squishlist}

We fix an intermediate threshold $\delta = 0.3$ to define the set of partisans and gatekeepers for each dataset.
We build balanced classification tasks by picking an equal number of partisans/gatekeepers and a random sample of non-partisan/non-gatekeeper users.

The accuracy of the classification model is shown in Table~\ref{tab:prediction} 
(average for 10-fold cross-validation) for partisans ($p$) and gatekeepers ($g$).
Given that the classification datasets are balanced, a random guess would have an accuracy of $0.5$.
However, all features give a better prediction.
It is interesting to see that just using simple n-gram features performs well.
This hints that there are marked differences in the way partisans and gatekeepers use text.
Note that n-gram features, even though using content, are not related to the production/consumption polarity computation, as these scores are only computed using tweets with links to news sources (and not the actual content itself).
Identifying partisans shows to be markedly easier than gatekeepers, 
with accuracies hovering around $80\%$ for partisans compared to $70\%$ for gatekeepers, 
when using all features combined.
Therefore, we conclude that being a partisan has clear correlations with specific network and content features that enable their identification with high accuracy.

\begin{table}[]
\centering
\caption{Accuracy for prediction of users who are partisans~($p$) or gatekeepers ($g$). (net) indicates network and profile features only, ($n$-gram) indicates just n-gram features. The last two columns show results for all features combined.}
\label{tab:prediction}
\begin{small}
\begin{tabular}{l c c c c c c}
\toprule
	& $p$ (net) & $g$ (net) & $p$ ($n$-gram) & $g$ ($n$-gram) & $p$ & $g$ \\
\midrule
\election   & 0.71      & 0.67  & 0.73 & 0.65 & 0.81 &  0.67    \\
\guncontrol & 0.70      & 0.64  & 0.76 & 0.62 & 0.83 &  0.67    \\
\obamacare  & 0.75      & 0.65  & 0.78 & 0.64 & 0.83 &  0.66    \\
\abortion   & 0.71      & 0.63  & 0.76 & 0.65 & 0.80 &  0.69   	\\
\largetw	& 0.72 		& 0.70 	& 0.74 & 0.68 & 0.78 & 0.75 	\\
\bottomrule
\end{tabular}
\end{small}
\vspace{-\baselineskip}
\end{table}

\section{Discussion}
\label{sec:discussion}


In this paper we study echo chambers in political discussions in social media, 
in particular, we study the interplay between content and network, 
and the different roles of users. 
Germane to our approach is the definition of measures for the political leaning of content 
shared by users in social media.
These measures, which are grounded in previous research~\citep{bakshy2015exposure},
capture both the leaning of the content shared by a single user, 
as well as the leaning of the content to which such user is exposed, 
by virtue of its neighborhood in the social network.

\spara{Characterising echo chambers}.
When applied to discussions about politically contentious topics,
our results support the existence of political echo chambers.
In particular, the distribution of production and consumption polarities of users is clearly bi-modal, 
and the production and consumption polarities are highly correlated.
Conversely, the phenomenon does not manifest itself when the topic of discussion is not contentious.
This result reinforces the validity of the proposed measures --- 
and agrees with similar conclusions presented by~\citet{barbera2015birds}, 
where retweet networks exhibit higher polarization for political topics.



\spara{Partisan users}.
We highlight the ``price of bipartisanship'' in terms of various aspects, 
including network position, community connections, and content endorsement.
Overall, bipartisan users pay a price in terms of network centrality, 
community connection, and endorsements from other users (retweets, favorites).
This is the first study to show the price of being bipartisan, 
especially in the context of political discussions forming echo chambers.
This result highlights a worrying aspect of echo chambers, 
as it suggests the existence of latent phenomena that effectively stifle mediation between the two sides.

\spara{Gatekeepers}.
Finally, we examined gatekeepers, i.e., users who are bipartisan consumers but partisan producers.
These users lie in-between the two opposed communities in network terms, but side with one in content terms.
Their clustering coefficient is usually lower, as they have links to both communities, which are unlikely to be closed.
%
The role of gatekeepers has not been examined in the context of echo chambers. Previous studies on Twitter showed that gatekeepers are typically ordinary citizens~\cite{xu2014talking} rather than officially active partisans (e.g., party members).

We also experimented with a different definition of gatekeepers -- users who have a high consumption variance and low production variance. 
This definition captures a slightly broader set of users (compared to Equation~\ref{eq:gatekeepers}),
e.g., users who consume from both ends of the political spectrum and produce balanced `centrist' content.
The results were almost identical to the ones reported above in Section~\ref{subsec:gatekeepers}, and so we do not present them explicitly.

Nevertheless, from our current analysis, it is not clear if such users act as open-minded net-citizens or ``sentinels'' who want to be informed about and attack the opinions of the opposition. 
Given the importance such users appear to have in the network structure 
(higher PageRank, and higher indegree (more followers)), 
this aspect remains to be studied in future work.
In the former case (i.e., if gatekeepers are open-minded net-citizens), 
gatekeepers would be good candidates for users to nudge towards the opposing 
side~\cite{garimella2017reducing,garimella2017nips,matakos2017measuring}.
The possibility of identifying gatekeepers to a non-random extent by just using network features 
(e.g., if they do not actively produce content) makes an interesting application.

\spara{Limitations}.
As with any empirical work, this study has its limitations.
First, the datasets used are just a sample of all the discussions in social media, 
and they all come from Twitter.
Twitter is, naturally, one of the main venues for online public discussion, 
and one of the few for which data is available -- hence a natural setting where to study echo chambers.
We tried to address concerns about the generality of our results by performing analysis on datasets of various sizes, from various domains and time periods.
However, as we focused on politically-savvy users on Twitter, the reader should not infer that our observations generalize immediately to other settings, or that echo chamber effects are as pronounced for the general public.

Second, our production and consumption scores rely on external labeling of news sources along a political axis.
This choice limits the applicability of our analysis to debates that are politically aligned, and 
mostly for English-speaking and US-related topics.
This limitation is not inherent in the methodology, 
but is due simply to the availability of such data.
Media bias and labeling of media on a political axis is a field in itself 
(media and communication studies), and hence, 
this is not a big limitation.
See the work by~\citet{groeling2013media} for a review on media bias and ways to label media sources.

Moreover, our analysis assumes that each user consumes all content produced by all their neighbors.
That is, we use the ``follow'' relationship as a proxy for the actual content consumption.
In reality, a user might not consume everything produced by the users they follow.
In the absence of scroll or click logs, which could give us more fine-grained results, 
this proxy is the best we can get.
%

Finally, it is possible that not all news articles from the news sources we base the polarity measures are political. During pre-processing, we attempted to split news articles from these sources into hard (politics, opinion, etc.) and soft news (gossip, entertainment, etc.) and applied the classifier from~\citet{bakshy2015exposure}. We found that almost all (over 85\%) of the urls from these domains are classified as hard news --- and so, we opted to consider all of them in our analysis, knowing that a small fraction of them might not be ``hard'' political news.


\spara{Future work}.
The results shown in this study are just one step towards the understanding of echo chambers and the interplay between network and content, which open up several directions for future work.

First, exploring more nuanced content and network features, which might lead to a better understanding of echo chambers in social media.
For instance, $n$-gram features turned out to be very informative for identifying partisans, which indicates a distinctive writing style of this set of users.
In this study we focused on content polarity based on a ground truth, but more powerful NLP techniques (e.g., topic modeling) might enable more powerful analysis.

Second, designing (probabilistic generative) models to capture the observed echo-chamber structure in terms of content and network features -- and the different user roles (i.e., (bi)partisan users and gatekeepers) and the price of bipartisanship.
Our findings show the interaction between network importance and the content produced and consumed by a user. 
Most of the existing models for dynamics of opinion formation and polarization on social networks either use exclusively content features, or use a dynamic process on a fixed random network~\cite{banisch2017opinion}.
However, in light of the current results, a comprehensive model for polarization should affect not only the opinion spread over the social network, but also the structure of the network itself.




\noindent
\textbf{Acknowledgments.}
This work was supported by the Academy of Finland projects ``Nestor'' (286211), ``Agra'' (313927), and ``AIDA'' (317085), and the EC H2020 RIA project ``SoBigData'' (654024).

\balance
\bibliographystyle{ACM-Reference-Format}
\bibliography{biblio}

\end{document}